\documentclass[12pt,aps,floats,showpacs,amssymb,tightenlines]{revtex4}

\UseRawInputEncoding
\usepackage{color}
\usepackage{graphicx}
\usepackage{amsmath}
\usepackage{amsfonts}
\usepackage{amscd}
\usepackage{epsfig}
\usepackage{amssymb}
\usepackage{tabularx}
\usepackage{float}

\begin{document}

\title{Linear perturbations of the Linet - Tian metrics with a
positive cosmological constant.}
\author{Reinaldo J. Gleiser} \email{gleiser@fis.uncor.edu}

\affiliation{Instituto de F\'{\i}sica Enrique Gaviola and FAMAF,
Universidad Nacional de C\'ordoba, Ciudad Universitaria, (5000)
C\'ordoba, Argentina}

\begin{abstract}

The Linet - Tian metrics are solutions of the Einstein equations
with a cosmological constant, $\Lambda$, that can be positive or
negative. The linear instability of these metrics in the case
$\Lambda <0$, has already been established. In the case $\Lambda>0$,
it was found in a recent analysis that the perturbation equations
admit unstable modes. The analysis was based on the construction of
a gauge invariant function of the metric perturbation coefficients,
called here $W(y)$. This function satisfied a linear second order
equation that could be used to set up a boundary value problem
determining the allowed, real or purely imaginary frequencies for
the perturbations. Nevertheless, the relation of these solutions to
the full spectrum of perturbations, and, therefore, to the evolution
of arbitrary perturbations, remained open. In this paper we consider
again the perturbations of the Linet - Tian metric with $\Lambda
>0$, and show, using a form of the Darboux transformation, that one
can associate with the perturbation equations a self adjoint problem
that provides a solution to the completeness and spectrum of the
perturbations. This is also used to construct the explicit relation
between the solutions of the gauge invariant equation for $W(y)$,
and the evolution of arbitrary initial data, thus solving the
problem that remained open in the previous study. Numerical methods
are then used to confirm the existence of unstable modes as a part
of the complete spectrum of the perturbations, thus establishing the
linear gravitational instability of the Linet - Tian metrics with
$\Lambda >0$.

\end{abstract}


\maketitle

\section{Introduction}

The Linet - Tian metrics \cite{linet}, \cite{tian}, are static
solutions of the Einstein equations with a cosmological constant,
$\Lambda$, that can be positive or negative.  They posses also two
other commuting Killing vectors: $\partial_{\phi}$, and
$\partial_{z}$ . The metrics are characterized by two constants: one
is $\kappa$, associated to the singularities of the metrics, and the
other is the cosmological constant $\Lambda$ \cite{conicity}. In the
limit of vanishing cosmological constant they reduce to a form of
the Levi - Civita metric \cite{levi}, and, therefore, they can be
considered as generalizations of the former to include a
cosmological constant. We refer to \cite{bronnikov} for a recent
review and bibliography on these and related types of metrics. Both
the Levi - Civita metric, and the Linet - Tian solution with
negative cosmological constant, have been found to be
gravitationally unstable, \cite{glei1}, \cite{glei2}. The positive
cosmological constant case was analyzed in \cite{glei3}, where it
was found that (linear) perturbations that break the symmetry
associated to $\partial_z$ (or $\partial _{\phi}$) contain unstable
modes. However, because of certain peculiar aspects of the
perturbation equations, the relation between those modes and the
general evolution of arbitrary perturbations remained unclear. The
purpose of this paper is to review and extend the results obtained
in \cite{glei3}, and to provide the proof that the existence of
unstable modes indeed implies that the positive $\Lambda$ Linet -
Tian space times are unstable.

The plan of the paper is as follows. In the next Section we consider
a particular form of the Linet - Tian metric that is suitable for
the analysis of the perturbation problem, and review and discuss
some of its properties, relevant for the present discussion. In
Section III, we consider the general linear perturbation of the
Linet - Tian metric, taking into account the existence of its
Killing vectors. In the present analysis we restrict the
perturbations to the ``diagonal'' case  considered in \cite{glei3},
and review the problems that the presence of gauge ambiguities pose
for setting up a meaningful perturbation analysis of the linear
stability of the metric. Some details regarding gauge
transformations are included for completeness in Appendix D. The
problem regarding the gauge ambiguities is solved in Section IV,
where we introduce a gauge invariant formulation, directly related
to a gauge invariant function ($W(y)$) of the perturbation
functions. This not only sidesteps the gauge ambiguities, but, as
shown, $W(y)$ provides a ``master function'', from which one may
compute the metric perturbation coefficients. This, is turn, makes
clear the reasons why a formulation directly in terms of the metric
perturbation functions will always be subject to gauge ambiguities.
Given this situation, it seems appropriate to look directly for the
equation satisfied by $W(y)$. This equation, that is derived in
Section V, has the form of a linear boundary value problem, with
eigenvalues $\Omega^2$, where $\Omega$ is the (real or pure
imaginary) frequency of the resulting perturbation modes. Using a
well known procedure, we obtain a related Schr\"odinger like
equation, but with a singular ``potential''. The singularity in the
``potential'' is related to singularities in the coefficients of the
equation satisfied by $W(y)$, although, as is shown in Appendix A,
$W(y)$ is regular at the singular point of the coefficients. In
Section VI, with some details given in Appendix B, and Appendix C,
we show that we can relate the analysis of completeness and spectrum
of the solutions for $W(y)$ to that of a self adjoint problem,
through a Darboux transformation. In Section VII we consider the
form that results from the previous analysis for the general
evolution of given perturbative initial data, making clear that
negative eigenvalues correspond to unstable modes. Numerical results
that confirm the existence of unstable modes are given in Section
VIII. In Section IX we consider the special cases where the
parameter $\kappa$ takes to values zero or one. The case of purely
``radial'' perturbations is analyzed in Section X. In Section XI we
discuss briefly the results obtained and the conclusions we extract
from them.

\section{The Linet - Tian metric with a positive cosmological
constant, and some of its properties.}

The Linet-Tian metrics with a positive cosmological constant may be
(locally) written as \cite{glei3},
\begin{eqnarray}\label{intro02}
     ds^2 & = & - y^{1/3+p_1/2} (1-\Lambda y )^{1/3-p_1/2} dt^2 +
     \frac{1}{3 y (1-\Lambda y)} dy^2 \nonumber \\ && +
y^{1/3+p_2/2} (1-\Lambda y )^{1/3-p_2/2} dz^2+ y^{1/3+p_3/2}
(1-\Lambda y )^{1/3-p_3/2} d\phi^2
\end{eqnarray}
where $\Lambda$ is the cosmological constant, and the parameters
$p_i$ are constrained by,
\begin{equation}\label{intro04}
  p_1+p_2+p_3=0\;\;;\;\; p_1^2+p_2^2+p_3^2 =\frac{8}{3}
\end{equation}
They may be given in terms of a single parameter $\kappa$
\cite{thorne}:
\begin{equation}\label{intro06}
    p_1=\frac{2(2 \kappa +2 \kappa^2-1)}{3 (1
    +\kappa+\kappa^2)}\;\;,\;\; p_2=
    -\frac{2(2 +4\kappa +\kappa^2)}{3 (1
    +\kappa+\kappa^2)}\;\;,\;\; p_3=
    \frac{2(2+2 \kappa - \kappa^2)}{3 (1
    +\kappa+\kappa^2)}
\end{equation}
In what follows we restrict $\kappa$ to the range $0 \leq \kappa
\leq 1$.

 We may naturally assume the range $-\infty < t < \infty$,
and $0 \leq y \leq 1/\Lambda$, but the ranges of $z$ and $\phi$
require further consideration \cite{wang} \cite{griffiths}. This is
because of the symmetrical roles played by $\partial_z$ and
$\partial_{\phi}$. This can be seen as follows. If we change
coordinates in (\ref{intro02}) as,
\begin{equation}\label{intro08}
y = \frac{1}{\Lambda} -x\;\;;\;\; t = \Lambda^{p_1/2}
\widetilde{t}\;\;;\;\; z = \Lambda^{p_2/2} \widetilde{\phi}\;\;;\;\;
\phi = \Lambda^{p_3/2} \widetilde{z}
\end{equation}
the metric takes the form,
\begin{eqnarray}\label{intro10}
     ds^2 & = & - x^{1/3-p_1/2}
     (1-\Lambda x )^{1/3+p_1/2} d\widetilde{t}^2 +
     \frac{1}{3 x (1-\Lambda x)} dx^2 \nonumber \\ && +
x^{1/3-p_2/2} (1-\Lambda x )^{1/3+p_2/2} d\widetilde{\phi}^2+
x^{1/3-p_3/2} (1-\Lambda x )^{1/3+p_3/2} d\widetilde{z}^2
\end{eqnarray}
If we now define $\eta$ such that,
\begin{equation}\label{intro12}
    \kappa=\frac{1-\eta}{2 \eta+1}
\end{equation}
and,
\begin{equation}\label{intro14}
    \widetilde{p}_1=\frac{2(2 \eta +2 \eta^2-1)}{3 (1
    +\eta+\eta^2)}\;\;,\;\; \widetilde{p}_2=
    -\frac{2(2 +4\eta +\eta^2)}{3 (1
    +\eta+\eta^2)}\;\;,\;\; \widetilde{p}_3=
    \frac{2(2+2 \eta - \eta^2)}{3 (1
    +\eta+\eta^2)}
\end{equation}
we find that (\ref{intro10}) can be written in the form,
\begin{eqnarray}\label{intro16}
     ds^2 & = & - x^{1/3+\widetilde{p}_1/2}
     (1-\Lambda x )^{1/3-\widetilde{p}_1/2} d\widetilde{t}^2 +
     \frac{1}{3 x (1-\Lambda x)} dx^2 \nonumber \\ && +
x^{1/3+\widetilde{p}_2/2} (1-\Lambda x )^{1/3-\widetilde{p}_2/2}
d\widetilde{z}^2+ x^{1/3+\widetilde{p}_3/2} (1-\Lambda x
)^{1/3-\widetilde{p}_3/2} d\widetilde{\phi}^2
\end{eqnarray}
which is identical to (\ref{intro02}), but with $\kappa$ replaced by
$\eta$ and an interchange of the roles of $\partial_z$ and
$\partial_{\phi}$. This implies that any Linet-Tian metric with a
given $\kappa$ and (positive) $\Lambda$ is locally isometric to a
Linet-Tian metric with the same $\Lambda$, but with $\kappa$
replaced by $(1-\kappa)/(2 \kappa+1)$. In particular, this implies
that if we find an instability for $\kappa$ in the range,
\begin{equation}\label{intro18}
  0 \leq  \kappa \leq (\sqrt{3}-1)/2 =0.366...
\end{equation}
then that instability will also be present for $(\sqrt{3}-1)/2 <
\kappa \leq 1$. Other implications of this symmetry have been
explored in the above mentioned references.

\section{Setting up the problem.}

Consider a general linear perturbation of the Linet - Tian metric.
This may be written in the form,
\begin{equation}\label{gau01}
     g_{\mu\nu}(t,y,z,\phi) =  g_{\mu\nu}^{(0)}(y)+\epsilon
      h_{\mu\nu}(t,y,z,\phi)
\end{equation}
where $g_{\mu\nu}^{(0)}(y)$ is the (unperturbed) Linet-Tian metric
(\ref{intro02}), and $\epsilon$ is an auxiliary parameter, used to
keep track of the linearity of the perturbations. The functions
$h_{\mu\nu}$ represent the most general perturbation. Since
$\partial_t$, $\partial_z$, and $\partial_{\phi}$ are commuting
Killing vectors of $g_{\mu\nu}^{(0)}$, we may restrict our analysis
of the perturbation equations to solutions of the form,
\begin{equation}\label{gau03}
    h_{\mu\nu}(t,y,z,\phi) = e^{i(\Omega t - k z -\ell \phi)}
    f_{\mu\nu} (y,\Omega,k,\ell)
\end{equation}

The evolution of a general perturbation would then be expressed as,
\begin{equation}\label{gau03a}
    h_{\mu\nu}(t,y,z,\phi) = \sum_{\Omega} \sum_{k} \sum_{ \ell}
    {\cal{C}}_{\Omega,k,\ell}\; e^{i(\Omega t - k z -\ell \phi)}
    f_{\mu\nu} (y,\Omega,k,\ell)
\end{equation}
where $\sum\limits_{a}$ stands for either a sum or an integral
depending on whether the corresponding variable is discrete or
continuous. Implicit in this expansion is the assumption that the
set of solutions $\{f_{\mu\nu} (y,\Omega,k,\ell)\}$ is ``complete''
in some appropriate sense, so that (\ref{gau03a}) is valid for a
general perturbation $h_{\mu\nu}(t,y,z,\phi)$. The expectation here
is that this could be achieved by imposing appropriate boundary
condition on the solutions of the perturbation equations. There are,
however, as indicated in \cite{glei3}, a number of difficulties in
establishing if, and in what sense, the expansion (\ref{gau03a}) has
the desired properties. One of these difficulties stems from the
fact that, as analyzed in \cite{glei3}, the general expansion
(\ref{gau01}) is subject to gauge ambiguities, and, therefore it may
contain components whose evolution in time is not determined by the
evolution equations. Some relevant details regarding this problem
are given in Appendix D.

In the present analysis, we consider again the ``diagonal'' case of
\cite{glei3}, where, for simplicity we take $\ell=0$, and the
perturbed metric is restricted to the form,
\begin{eqnarray}
\label{gau07}
  ds^2 &=& -\frac{y^{\frac{1}{3}+\frac{p_1}{2}}
  \left(1+\epsilon e^{i(\Omega t
  - k z)}F_1(y)\right)}{(1-\Lambda y)^{\frac{p_1}{2}-\frac{1}{3}}}  dt^2
        +\frac{\left(1+\epsilon e^{i(\Omega t - k z)}F_2(y)\right)}
        {3  y(1-\Lambda y)}  dy^2  \nonumber \\
   & & +\frac{y^{\frac{1}{3}+\frac{p_2}{2}}\left(1+\epsilon
   e^{i(\Omega t - k z)}F_3(y)\right)}{(1-\Lambda y)^{\frac{p_2}{2}
   -\frac{1}{3}}}  dz^2
   +\frac{y^{\frac{1}{3}+\frac{p_3}{2}}\left(1+\epsilon
   e^{i(\Omega t - k z)}F_4(y)\right)}
   {(1-\Lambda y)^{\frac{p_3}{2}-\frac{1}{3}}}  d\phi^2  \;,
\end{eqnarray}

This choice is consistent with the equations of motion but, as
indicated in Appendix D, and analyzed below, it is not free from
gauge ambiguities.

Consider now the linearized Einstein equations for the metric
(\ref{gau07}). For $\Omega \neq 0$, and $k\neq 0$, we have,
\begin{equation}
\label{diag03}
 F_2(y)=-F_4(y),
\end{equation}
and the remaining equations can be written in the form,
\begin{equation}
 \label{diag04a}
  \frac{dF_1}{dy} +\frac{dF_4}{dy} +\frac{p_1-p_2}{4 y
  (1-\Lambda y)} F_1
   -\frac{8 \Lambda y+9 p_2 -4 +3 p_1}{12 y (1-\Lambda y)} F_4  =0,
\end{equation}
\begin{equation}
 \label{diag04b}
  \frac{dF_3}{dy} +\frac{dF_4}{dy} -\frac{p_1-p_2}{4 y
  (1- \Lambda y)} F_3
   -\frac{8 \Lambda  y+9 p_1 -4 +3 p_2}{12 y (1-\Lambda y)} F_4 =0,
\end{equation}
and,
\begin{eqnarray}
\label{diag06}
 {\frac {d F_4}{dy}} & = & - {\frac {2
\left( 1-\Lambda y \right) ^{-1/3+ p_1/2} \left( F_{{3}} +F_{ {4}}
\right) {\Omega}^{2}}{{y}^{1/3+p_1/2}
 \left( -2-3 p_3+4\,\Lambda\,y \right) }}+{\frac {2 \left( 1-
\Lambda\,y \right) ^{-1/3+p_2/2} \left( F_{{4}}  +F_{{1}}  \right)
{k}^{2}}{{y}^{1/3+p_2/2} \left( -2-3 p_3+4 \Lambda y \right) }}
\nonumber \\ & & +{\frac {
 \left( 8\,\Lambda\,y+3\,p_1-4 \right)  \left( p_1-p_2
 \right) F_{{1}}  }{8 y \left( -1+\Lambda\,y \right)
 \left( -2-3\,p_3+4\,\Lambda\,y \right) }}
\nonumber \\ & &  -{\frac { \left( 8 \Lambda \left( p_1-p_2 \right)
y+ \left( 2+3 p_1
 \right)  \left( p_1+2\,p_2-2 \right)  \right) F_{{3}}
  }{8 y \left( -1+\Lambda\,y \right)
   \left( -2-3 p_3+4 \Lambda y \right) }}
  \nonumber \\ & &
-{\frac { \left( 32\,{\Lambda}^{2}{y }^{2}-8 \Lambda  \left( 4+15
p_3 \right) y+60 p_3+45 p_1 p_2+44 \right) F_{{4}} }{24 y \left( -1+
\Lambda y \right)  \left( -2-3 p_3 +4 \Lambda y \right) }}
\end{eqnarray}

Since $4\Lambda y-3p_3-2 = 4 \Lambda
y-6(\kappa+1)/(1+\kappa+\kappa^2) < 0$ in $0 < y < 1/\Lambda$, the
$y$-dependent coefficients of the $F_i$ are regular in $0 < y
<1/\Lambda$, but singular, in general, for both $y=0$ and
$y=1/\Lambda$. Therefore, the general solution of the system
(\ref{diag04a})-(\ref{diag04b})-(\ref{diag06}) can be written as a
linear combination of {\em three} appropriately chosen linearly
independent solutions, which are {\em regular}, in $0 < y <1/\Lambda
$, but may be singular at either or both $y=0$, and $y=1/\Lambda$.
In particular, the set (see Appendix D and \cite{glei3}),
\begin{eqnarray}
\label{diag08}
  F_1(y) &=& \frac{(3p_1+2-4 \Lambda y)y^{p_1/4+p_2/4-2/3}(p_1-p_2) }
          {(1- \Lambda y)^{p_1/4+p_2/4+2/3}} +\frac{16 y^{p_2/4-p_1/4}
           \Omega^2
          }{(1-\Lambda y)^{p_2/4-p_1/4}} \nonumber \\
  F_3(y) &=&  \frac{(3p_2+2-4 \Lambda y)y^{p_1/4+p_2/4-2/3}(p_1-p_2) }
          {(1-\Lambda y)^{p_1/4+p_2/4+2/3}} +\frac{16 y^{p_1/4-p_2/4} k^2
          }{(1- \Lambda y)^{p_1/4-p_2/4}} \nonumber \\
    F_4(y) &=& \frac{(2-4 \Lambda y+3p_3)(p_1-p_2)}
    { y^{2/3+p_3/4}(1- \Lambda y)^{2/3-p_3/4}}
\end{eqnarray}
is an exact, but pure gauge solution of the system
(\ref{diag04a})-(\ref{diag04b})-(\ref{diag06}), that can be removed
by an appropriate coordinate transformation. Notice that, as
indicated, this solution is regular for $0 < y < 1/\Lambda$, but it
is divergent both for $y \to 0$ and $y \to 1/\Lambda$.\\

As regards the other two independent solutions, we have that one of
them, near $y=0$, behaves as,
\begin{eqnarray}
\label{diag10}
  F_1(y) & \simeq &  -\frac{2+4\kappa+\kappa^2}{\kappa(2+\kappa)}
   c_0 + a_1 y^{\frac{1}{1+\kappa+\kappa^2}} \nonumber \\
  F_3(y) & \simeq &  -\frac{\kappa^2-2}{\kappa(2+\kappa)} c_0
  + b_1 y^{\frac{1}{1+\kappa+\kappa^2}} \nonumber \\
    F_4(y) & \simeq &   c_0 + c_1 y^{\frac{1}{1+\kappa+\kappa^2}}
\end{eqnarray}
plus higher order terms, where $c_0$ is an arbitrary constant, and
\begin{eqnarray}
\label{diag12}
  a_1 & = &  \frac{(2+\kappa)(\kappa^2-2+2\kappa)(1+\kappa+\kappa^2)^2
   \Omega^2 c_0}{3\kappa(\kappa^2+2\kappa+4)} \nonumber \\
   b_1 & = &  \frac{(\kappa-2)(\kappa^2+2+2\kappa)(1+\kappa+\kappa^2)^2
   \Omega^2 c_0}{3\kappa(\kappa^2+2\kappa+4)} \nonumber \\
     c_1 & = &  \frac{(2-2\kappa-\kappa^2)(\kappa^2+2+2\kappa)
     (1+\kappa+\kappa^2)^2\Omega^2 c_0}{3(2+\kappa)(\kappa^2+2\kappa+4)}
\end{eqnarray}
and, therefore, the $F_i$ approach a finite limit as $y \to 0$, but
with divergent derivatives in that limit, because
$(1+\kappa+\kappa^2)^{-1} < 1$, for $\kappa >0$.

For the other solution, near $y=0$, we have,
\begin{eqnarray}
\label{diag14}
  F_1(y) & \simeq &  -\frac{2+4\kappa+\kappa^2}{\kappa(2+\kappa)}c_2
   \ln(y)+ \frac{4(1+\kappa)(1+\kappa+\kappa^2)}{\kappa^2(2+\kappa)^2}c_2
   -\frac{2+4\kappa+\kappa^2}{\kappa(2+\kappa)}c_3 \nonumber \\
  F_3(y) & \simeq &  \frac{2-\kappa^2}{\kappa(2+\kappa)}c_2
  \ln(y)+ \frac{4(1+\kappa)(1+\kappa+\kappa^2)}{\kappa^2(1+\kappa)^2}
   c_2 -\frac{\kappa^2-2}{\kappa(2+\kappa)}c_3  \nonumber \\
    F_4(y) & \simeq &   c_2
    \ln(y)+c_3
\end{eqnarray}
where $c_2$ and $c_3$ are an arbitrary constant, plus terms that
vanish as $y \to 0$, and, therefore, the $F_i$ diverge as $\ln(y)$.
The presence of an additional constant, $c_3$, in these expressions
is due to the fact that to any solution satisfying the boundary
condition (\ref{diag14}) we may add an arbitrary solution satisfying
(\ref{diag10}) without changing the form (\ref{diag14}).

Similarly, near $y=1/\Lambda$, we have a solution that behaves as,
\begin{eqnarray}
\label{diag16}
  F_{{1}} \left( y \right) &=& {\frac { \left(\mu -9\kappa-3 \right)
   c_4}{3\kappa \left( 2+\kappa \right) }}
  +\frac{3\kappa
  \left(\kappa -4 \right) c_5 }  {\left(
9 \kappa+\mu-3
 \right)} \left( 1-\Lambda y \right)
^{{ \frac { \left(1 -\kappa \right) ^{2}}{3\mu}}}
 \nonumber \\
 F_{{3}} \left( y \right) &=&{\frac { \left(
9-7 \mu+3 \kappa \right) c_4 }{3 \kappa \left( 2+\kappa \right)
}}+\frac{9
 \left( \mu-1-\kappa \right) c_5}  { \left(
3+3\kappa-5 \mu \right)} \left( 1-\Lambda y \right) ^{{\frac
{ \left(1 -\kappa \right) ^{2}}{3\mu}}} \nonumber \\
 F_{{4}} \left( y
\right) & = & c_{{4}}+c_{{5}} \left( 1 -\Lambda y \right) ^{{\frac {
\left( -\kappa+1 \right) ^{2}}{3\mu }}}
\end{eqnarray}
plus higher order terms, $c_4$ is an arbitrary constant,
$\mu=1+\kappa+\kappa^2$, and,
\begin{equation}\label{diag16a}
 c_5=\frac{ \left( 9 \kappa+\mu-3 \right)  \left(5 \mu  -3-3
\kappa \right) {\mu}^{2} k^2 c_4}{  {\Lambda}^{{\frac {\kappa}{\mu}
}+1} {\kappa} \left( 2+\kappa \right) \left( 7 \mu-3-9\kappa \right)
\left( \kappa-1 \right) ^{4}}
\end{equation}

For the other independent solution $F_1$, $F_3$ and $F_4$ diverge as
$\ln(1-\Lambda y)$ as $y \to 1/\Lambda$, but we shall not display
their leading behaviour for simplicity. Thus, we see that the system
has solutions that are well behaved, i.e., do not diverge, at either
$y=0$ or $y=1/\Lambda$.\\

What this means is that if we consider a solution that behaves as
(\ref{diag10}) near $y=0$, then, in general, as we approach
$y=1/\Lambda$, it will behave as a linear combination of the three
linearly independent solutions characterized by their behaviour near
$y=1/\Lambda$, and, therefore, it will diverge for $y \to
1/\Lambda$. As discussed, for instance in \cite{glei1} or
\cite{glei2}, we should, in principle, consider only as appropriate
those solutions of the perturbation equations such that the $F_i$,
(up to a gauge transformation) do not diverge either at $y=0$ or
$y=1/\Lambda$. Since solutions of the system can only be obtained
numerically, one might then try to impose this condition at say
$y=0$, and, for fixed $\kappa$ and $k$, look for possible values of
$\Omega$, such that the solution is also finite as we approach
$y=1/\Lambda$. Unfortunately, because of the gauge ambiguities
contained in the system, this simple ``shooting'' procedure fails to
provide the required solutions. What is required here is a {\em
gauge invariant function} that carries the physical properties of
the perturbations, and satisfies the finiteness requirements, while
the $F_i$ themselves may still contain gauge dependent divergent
components. This problem is considered in the next Section.

\section{Gauge invariant formulation.}

Gauge invariant functions may be constructed in general as a linear
combinations of the $F_i(y)$. Let us call $F^{g}_i(y)$ the solutions
given by (\ref{diag08}). Then, a suitable example is the function,
\begin{equation}\label{diag18}
    W(y) =
    {\cal{K}}(y)\left[F^{g}_3(y)F_4(y)-F^{g}_4(y)F_3(y)\right]
\end{equation}
where ${\cal{K}}$ is an arbitrary function of $y$. If we choose,
\begin{equation}\label{diag20}
 {\cal{K}}( y ) =
 -\frac{{\mu}^{2}}{4 }{y}^{{\frac {1+\kappa+\mu}{2\mu}}}
 \left( 1-\Lambda y \right) ^{\frac {5\mu-3-3
\kappa}{6\mu}},
\end{equation}
where here, and in what follows, $\mu=1+\kappa+\kappa^2$, we get,
\begin{eqnarray}
\label{diag22}
 W \left( y \right) & = &- \left(
2\Lambda y\mu-3-3\kappa
 \right) \kappa \left( 2+\kappa \right) F_{{3}} \left( y \right) \\
 && +
 \left(\kappa \left( 2+\kappa \right) \left(
3 \kappa +2 \Lambda y\mu \right) -4 \left( 1-\Lambda y \right)
^{{\frac { \left( \kappa -1 \right) ^{2}}{3\mu}}}{y}^{{\frac {
\left( 1+\kappa \right) ^{2}}{\mu }}}{\mu}^{2}{k}^{2}   \right)
F_{{4}} \left( y \right)\nonumber
\end{eqnarray}

Notice that the coefficients of $F_3$, and $F_4$ are finite both for
$y \to 0$ and $y \to 1/\Lambda$. In particular, near $y=0$, for the
solution (\ref{diag10}) we have,
\begin{equation}
\label{diag26}
 W(y) \simeq
  3(1 +\kappa+\mu) c_0 - (1 +\kappa+\mu) \mu^2 \Omega^2
   c_0 y^{\frac{1}{\mu}},
\end{equation}
and, near $y=1/\Lambda$, for the solution (\ref{diag16}) we have,
\begin{eqnarray}
\label{diag28}
 W\left( y \right) & = & \left(\frac{ 4}{3}\mu  \left(
5\,\mu-6 \right) \Lambda-4\,\kappa\,\mu+3\,\kappa-\mu+3 \right) c_4
\nonumber \\ & & + \frac{  \left( 4 \mu\, \left( 7 \mu-6-6 \kappa
\right) \Lambda-3  \kappa  \left( 2+\kappa \right)  \left( 4
\kappa-1 \right)  \right) {k}^{2 } {\mu}^{2} c_4}{  {\Lambda}^{{
\frac {\kappa}{\mu}}+1}  \left( 7 \mu-3-9 \kappa
 \right)  \left( \kappa-1 \right) ^{4}\left(
9\kappa+\mu-3 \right)^{-1}} \left( 1- \Lambda y \right) ^{{\frac {
\left(1 -\kappa \right) ^{2}}{3\mu} }}
\end{eqnarray}
plus higher order terms. Thus, $W$ will be well defined and finite
for solutions $F_3$, and $F_4$ that satisfy, up to an arbitrary
addition of the pure gauge solution, both the finite boundary
conditions (\ref{diag10}), and (\ref{diag16}). Then, the finiteness
of $W$ corresponds precisely to the condition for the existence of
appropriate solutions of the set (\ref{diag04a},\ref{diag04b},
\ref{diag06}).  \\

But the crucial property of $W$ is that it is not only gauge
invariant, but it is also a {\em master function}, in the sense that
the full perturbation can be reconstructed from $W$. This can be
seen as follows. First, we solve (\ref{diag22}) for $F_3$ in terms
of $W$, and, $F_4$,
\begin{equation}
\label{diag30}
  F_3(y) = \frac{ {\cal{K}}  F^{g}_3 F_4 -W}{ {\cal{K}}F^{g}_4}
\end{equation}

Replacing (\ref{diag30}) in (\ref{diag04b}), using the fact that the
$ F^{g}_i$ are solutions of (\ref{diag04b}), and rearranging terms
we find,
\begin{eqnarray}
\label{diag31}
  {\frac {d}{dy}} \left( \frac {F_4  }{F^{g}_4}
  \right)& = &{\frac {{W}
 \left( \kappa+\mu-1 \right) }{2 \mu F^{g}_4 {\cal{K}}
   \left( F^{g}_3 + F^{g}_4
   \right) y \left( 1-\Lambda y \right) }}+
  \frac{1}{F^{g}_3 + F^{g}_4}
   {\frac {d}{dy}}
 \left( {\frac {{W} }{F^{g}_4 {\cal{K}}  }} \right),
\end{eqnarray}
which implies,
\begin{eqnarray}
\label{diag32}
  F_4(y)& = & F^{g}_4 \int_0^y{\left[{\frac {
 \left( \kappa+\mu-1 \right) {W}}{2 \mu F^{g}_4 {\cal{K}}
   \left( F^{g}_3 + F^{g}_4
   \right) y' \left( 1- \Lambda y' \right) }}+
  \frac{1}{(F^{g}_3 + F^{g}_4)}
   {\frac {d}{dy'}}
 \left( {\frac {{W} }{F^{g}_4 {\cal{K}}  }} \right)\right] dy'}
 \nonumber \\
 && + C F^{g}_4(y),
\end{eqnarray}
where $C$ is an arbitrary constant, and, therefore, we can express
$F_4$ entirely in terms of $W$, and the already known pure gauge
solutions.

Using the expressions for $F_3$, and $F_4$ we may also obtain an
expression for $F_1(y)$, in terms of $W(y)$, but it turned out to be
more useful for the derivations to solve (\ref{diag06}) for
$F_1(y)$. This is given by,
\begin{eqnarray}
\label{diag32a}
 F_{{1}} & = & {\frac {4y \left(
1-\Lambda y \right)
 \left( 2 \Lambda y\mu-3-3 \kappa \right) \mu   }{A_{{1}}}}{\frac {d F_4}{dy}}
 -{\frac {A_{{2}}  }{3A_{{1}}}} F_{{4}}  \\
 & &+ \frac{ 4 \left( 1-\Lambda y \right) ^{{
\frac {4\mu-3}{3\mu}}}{y}^{\frac{1}{\mu}}{\Omega}^{2}{\mu}^{2}-4\mu
\Lambda \left( \mu+\kappa -1 \right) y+3 \left( \mu -1\right)
 \left( \mu+1+2\kappa \right) } {A_1} F_3 \nonumber
\end{eqnarray}
where,
\begin{equation}\label{diagA1}
 A_{{1}}=4 \left( 1-\Lambda y \right) ^{{\frac
 {\mu-3 \kappa}{3
\mu}}}{y}^{{\frac {\kappa+\mu}{\mu}}}{k}^{2}{\mu}^{2}- \left( 4
\Lambda y\mu-3 \right)  \left( \mu+\kappa-1 \right)
\end{equation}
and,
\begin{eqnarray}
\label{diagA2}
 A_2 & = &-12 \left( 1-\Lambda y \right) ^{{\frac {4\,\mu-3}{3
\mu}}}{y}^{\frac{1}{\mu}}{\Omega}^{2}{\mu}^{2}+12 \left( 1-\Lambda y
 \right) ^{{\frac {\mu-3\,\kappa}{3\mu}}}{y}^{{\frac {\kappa+\mu}{
\mu}}}{k}^{2}{\mu}^{2} \\ &&
 +8 {\Lambda}^{2}{\mu}^{2}{y}^{2}+12 \mu
\Lambda  \left(\mu -5-5 \kappa \right) y-9 {\mu}^{2}+45\kappa+45 \mu
\nonumber
\end{eqnarray}

Thus, the full set of diagonal perturbations  can be written in
terms of the master function $W$. The resulting expressions,
nevertheless, still contain gauge ambiguities. In fact, going back
to (\ref{diag32}), we can see as expected, that $F_4$ reduces to
$F^{g}_4$ when $W(y)=0$, the pure gauge situation. But, suppose now
that we insert in (\ref{diag32}) a non trivial $W(y)$, satisfying
the boundary condition (\ref{diag26}). It is easy to check that if
we also set $C=0$, the resulting $F_4(y)$ satisfies (\ref{diag10})
near $y=0$. But, we can also check that near $y=1/\Lambda$, since,
in general, the integral is finite, $F_4(y)$ approaches, in general,
$F^{g}_4(y)$. There is no contradiction here. It simply means that
we cannot choose a simple gauge where $F_4$ is free of $F^{g}_4(y)$
``contamination''. This suggests that we look directly for the
equation that $W(y)$ should satisfy, when the $F_i$ satisfy their
corresponding equations. This is derived in the next Section.

\section{The differential equation for $W(y)$.}

We may obtain the equation that $W(y)$ should satisfy if the
$F_i(y)$ satisfy the perturbation equations by going back to
(\ref{diag31}), and taking a new $y$-derivative. Solving for $d^2
W/dy^2$, and after several replacements, using the evolution
equations for the $F_i$, we finally find that $W$ satisfies the
equation,
\begin{equation}
\label{diag40}
  -{\frac {d^{2}W}{d{y}^{2}}} +Q_1{\frac {dW}{dy}} +Q_2 W
  ={\Omega}^{2} Q_3 W
\end{equation}
where,
\begin{eqnarray}
\label{diag40a}
  Q_1( y ) =\frac{
4{k}^{2}\mu \left( 2 \Lambda y\mu-3\mu-6\kappa \right) \left(
1-\Lambda y \right) ^{ {\frac { \left( \kappa-1 \right)
^{2}}{3\mu}}}{y}^{{\frac {
 \left( 1+\kappa \right) ^{2}}{\mu}}}+3 \left( 2+\kappa \right)
\kappa \left( 4\Lambda y\mu-6\Lambda y+3 \right) }{
 3 \left( 4 \left( 1-\Lambda y \right) ^{{\frac { \left( \kappa-
1 \right) ^{2}}{3\mu}}}{y}^{{\frac { \left( 1+\kappa \right)
^{2}}{\mu} }}{\mu}^{2}{k}^{2}-\kappa\, \left( 2+\kappa \right)
\left( 4\,\Lambda \,y\mu-3 \right)  \right) {y} \left(\Lambda y-1
\right) },
\end{eqnarray}

\begin{eqnarray}
\label{diag40b}
  Q_2( y )& = &
 \left[ 4{k}^{4}{\mu}^{2} \left( 1-\Lambda y \right)
^{-{\frac {6\kappa+\mu}{3\mu}}}{y}^{{\frac { 2\kappa+\mu}{\mu}}}
\right. \nonumber \\ & & \left. -{k}^{2} \left(
8{\Lambda}^{2}{\mu}^{2}{y}^{2} -4\Lambda \mu \left(3\kappa+5\mu
-3\right) y+3\kappa
 \left( 2\kappa+3 \right)  \left( 1+2\kappa \right)  \left( 2+
\kappa \right)  \right) \right. \nonumber \\ & & \left. \left(
1-\Lambda y \right) ^{-{\frac {3 \kappa+2\mu}{3\mu}}}{y}^{{\frac
{\kappa}{\mu}}}+6\Lambda\kappa \left( 2+\kappa \right) \left(
-3+2\mu \right)  \right]   \\
& &\left[ 3 {y}
 \left( 4\, \left( 1-\Lambda\,y \right) ^{{\frac { \left( \kappa-
1 \right) ^{2}}{3\mu}}}{y}^{{\frac { \left( 1+\kappa \right)
^{2}}{\mu} }}{\mu}^{2}{k}^{2}-\kappa\, \left( 2+\kappa \right)
\left( 4\Lambda y\mu-3 \right)  \right)  \left( 1-\Lambda y
\right)\right]^{-1}  \nonumber ,
\end{eqnarray}
and,
\begin{equation}
\label{diag40c}
 Q_3 \left( y \right) =\frac{\left( 1-\Lambda y
\right) ^{-{ \frac {3+2\,\mu}{3\mu}}}{y}^{{\frac {1-2\mu}{\mu}}}}{3}
\end{equation}

Thus, imposing appropriate boundary conditions on $W(y)$, we may
consider (\ref{diag40}) as a boundary value problem whose solutions
determine the allowed values of $\Omega$. Independently of the
details, to analyze an equation of the form (\ref{diag40}) it may be
useful to put it into a Schr\"odinger-like form, that possibly leads
to an equivalent self adjoint problem. This can be achieved
introducing a new coordinate $r=r(y)$, and two new functions,
$K(y)$, and $\widetilde{W}(r)$, such that,
\begin{equation}\label{diag44}
    W(y)=K(y)\widetilde{W}\left(r(y)\right)
\end{equation}

Replacing in (\ref{diag40}) we get,
\begin{eqnarray}\label{diag46}
& &-{\frac {d^{2} \widetilde{W}}{d{r}^{2}}} -{\frac { \left( 2
{\dfrac {dK}{dy}} {\dfrac {dr}{dy}}
  +K  {\dfrac {d^{2}r}{d{y}^{2}}} -{ Q_1} K  {
\dfrac {dr}{dy}}  \right)   }{K   \left( {\dfrac {dr}{dy}}
   \right) ^{2}}}{\frac {d \widetilde{W}}{dr}}-{\frac {
    \left( {\dfrac {d^{2}K}{d{y}^{2}}} -{Q_1}
  {\dfrac {d K}{dy}} -{Q_2}  K   \right) }{K   \left( {
\dfrac {d r}{dy}}  \right) ^{2}}}\widetilde{W} \nonumber \\
&&= \Omega^2 \frac{Q_3}{\left(\dfrac {dr}{dy} \right)^2}
\widetilde{W}
\end{eqnarray}

If we impose now that $r(y)$ be a solution of,
\begin{equation}
\label{diag48}
{\frac {d r}{dy}} =\sqrt{Q_3(y)},
\end{equation}
and also that $K(y)$ is a solution of,
\begin{equation}
\label{diag50}
2\dfrac {dr}{dy}{\frac {d K}{dy}} +
 \left( {\frac {d^{2}r}{d{y}^{2}}} -{ Q_1} {\frac {dr}{dy}}
  \right)K = 0,
\end{equation}
replacing in (\ref{diag46}), we find that $\widetilde{W}$ satisfies
the Schr\"odinger - like equation,
\begin{equation}
\label{diag52}
 -{\frac {d^{2} \widetilde{W}}{d{r}^{2}}} +
\mathbf{V}\widetilde{W}= \frac{\Omega^2}{\Lambda}\widetilde{W},
\end{equation}
where the ``potential'' $\mathbf{V}$ is given by,
\begin{equation}
\label{diag54}
\mathbf{V}  =\frac{1}{4 Q_3^3} \left(Q_3
    \frac{d^2Q_3}{dy^2}-
    \frac{5}{4}\left(\frac{dQ_3}{dy}\right)^2\right)
    +\frac{1}{4Q_3}\left(Q_1^2+4Q_2-2 \frac{dQ_1}{dy}\right)
\end{equation}
and, therefore, it is explicitly given as a function of $y$, even if
we do not have explicit solutions for either (\ref{diag48}) or
(\ref{diag50}). Actually, in our case $Q_3$ is given by
(\ref{diag40c}), and the general solution of (\ref{diag48}) is,
\begin{equation}\label{diag56}
r \left( y \right) = \frac{2 \mu {y}^{ \frac{1} {2\mu
 }}}{\sqrt {3}}\;
{{}_2F_1\left({\tfrac{1}{ 2\mu}} ,{\tfrac {3+2\mu}{6\mu}};{\tfrac
{1+2\mu}{2\mu}};{\Lambda y}\right)} +C_0,
\end{equation}
where ${}_2F_1(a,b;c;{x})$ is a hypergeometric function, with $C_0$
an arbitrary constant. In what follows, without loss of generality,
we will set $C_0=0$. The range of $r$ will be then,
\begin{equation}\label{diag56a}
    0 \leq r \leq r_1
\end{equation}
where $r_1$ is given by,
\begin{eqnarray}\label{diag56b}
    r_1 &= & \int_0^{1/\Lambda} \frac{1}{\sqrt{3}
     \left( 1-\Lambda y
 \right) ^{\frac{2\mu+3}{6\mu}}{y}^{\frac {2\mu-1}{2\mu}}} dy
 \nonumber \\
 &=& \frac{\Gamma\left(\frac{4 \mu -3}{6\mu}\right)\Gamma
 \left(\frac{1}{2\mu}\right)}{\sqrt{3}\Lambda^{\frac{1}
 {\mu}}\Gamma(2/3)}
\end{eqnarray}

We may use (\ref{diag56}) to construct a parametric representation
of $\mathbf{V}(r)$. This would in principle allow us, as in similar
quantum  mechanical problems, to carry out  a qualitative analysis
of the possible spectrum of allowed values of the ``eigenvalues''
$\Omega^2$, and therefore obtain information on the existence of
solutions with $\Omega^2 <0$, signalling unstable solutions of the
evolution equations. Unfortunately, in our case, irrespective of the
value of $k$, the functions $Q_1$, and $Q_2$ have vanishing
denominators at $y=y_0$, where $y_0$ is a solution of,
\begin{equation}
 \label{com02a}
 {k}^{2}=\frac{\kappa \left( \kappa+2 \right)
 \left( 4 \Lambda {y_0}\mu-3 \right)}
 {  4 \left( 1-\Lambda { y_0} \right)
^{{\frac { \left( \kappa-1 \right) ^{2}}{3\mu}}}
 {{y_0}}^{{\frac { \left( 1+\kappa \right) ^{2}}{\mu}}}
 {\mu}^{2}}
\end{equation}
Notice that, for $k^2> 0$, this equation has always a solution for
$y_0$ in the range $3/(4\Lambda \mu) \leq y_0 <1/\Lambda$, and, as
can be checked, this implies that $\mathbf{V}(y)$ has a double pole
at $y=y_0$, and, therefore, (\ref{diag52}) is not self adjoint, and
the analysis fails.

One may ask whether this problem could be solved with a different
choice of ${\cal{K}}(y)$ in (\ref{diag18}), that would lead to a
different equation for the resulting $W(y)$, and therefore, possibly
to a different associated Schr\"odinger like equation. That this is
not the case can be seen as follows. Suppose we introduce a new
function ${\cal{K}}_1(y)$, and define $W_1(y)$ by,
\begin{equation}\label{teo01}
    W(y)={\cal{K}}_1(y) W_1(y)
\end{equation}

Replacing in (\ref{diag18}) we get,
\begin{equation}\label{teo02}
-\frac{d^2 W_1}{dy^2}+ \widetilde{Q}_1 \frac{dW_1}{dy}
+\widetilde{Q}_2 W_1 =\Omega^2 \widetilde{Q}_3 W_1
\end{equation}
where,
\begin{eqnarray}
 \label{teo03}
  \widetilde{Q}_1(y) &=& Q_1-\frac{2}{{\cal{K}}_1}\frac{d {\cal{K}}_1}{dy} \nonumber \\
  \widetilde{Q}_2(y) &=& Q_2+\frac{Q_1}{{\cal{K}}_1}-\frac{d^2 {\cal{K}}_1}{dy^2} \\
  \widetilde{Q}_3(y) &=& Q_3 \nonumber
\end{eqnarray}

Then, defining $W_1(y) = K_1(y) \widetilde{W}_1(r(y))$, the same
procedure that led to (\ref{diag52}), leads now to,
\begin{equation}
\label{diag52a}
 -{\frac {d^{2} \widetilde{W}_1}{d{r}^{2}}} +
\mathbf{V}_1\widetilde{W}_1=
\frac{\Omega^2}{\Lambda}\widetilde{W}_1,
\end{equation}
where the ``potential'' $\mathbf{V}_1$ is now given by,
\begin{equation}
\label{diag54a}
 \mathbf{V}_1  =\frac{1}{4 \widetilde{Q}_3^3}
\left(\widetilde{Q}_3
    \frac{d^2\widetilde{Q}_3}{dy^2}-
    \frac{5}{4}\left(\frac{d\widetilde{Q}_3}{dy}\right)^2\right)
    +\frac{1}{4\widetilde{Q}_3}\left(\widetilde{Q}_1^2
    +4\widetilde{Q}_2-2 \frac{d\widetilde{Q}_1}{dy}\right)
\end{equation}
and, since $\widetilde{Q}_3(y) = Q_3(y)$, we have that $r(y)$ is the
same as in (\ref{diag52}). But if we replace now (\ref{teo03}) in
(\ref{diag54a}), we immediately obtain that $\mathbf{V}_1(y)\equiv
\mathbf{V}(y)$, and therefore, (\ref{diag52}) is invariant under a
change ${\cal{K}}(y)$, and the problem cannot be solved by a
different choice of this factor.\\

At this point we must remark that although some coefficients in
(\ref{diag40}) are singular, the solutions $W(y)$ must be regular at
$y_0$, because they are linear combinations of $F_3$ and $F_4$, with
regular coefficients, with $F_3$ and $F_4$ also regular in a
neighbourhood of $y=y_0$. In fact one can check, by explicit
computation, that near $y=y_0$, the general solution of
(\ref{diag40}) takes the form,
\begin{eqnarray}
\label{com02b}
  W &=&a_{{0}}+{\frac {a_{{0}} \left( 2\,\Lambda\,y_0\,\mu
  -2-\kappa \right)  \left( 2\,\Lambda\,y_0\,\mu-3\,\mu+3 \right) }
  {2 \left(
2\,\Lambda\,y_0\,\mu-3\,\kappa-3 \right)  \left( \Lambda\,y_0-1
\right) y_0\,\mu}} (y-y_0)
  \nonumber
\\ && +  a_{{0}}\left[
\frac{y_0^{\frac{1-2\mu}{\mu}}\Omega^2}{6(1-\Lambda
y_0)^{\frac{3+2\mu}{3 \mu}}} -{\frac { \left( 2+\kappa \right)
\left( 4\Lambda y_0 \mu-3 \right) \left( 2 \Lambda y_0 \mu-6\mu+9+3
\kappa
 \right)\kappa}{24 \left( 2\Lambda y_0\mu-3 \kappa-3
 \right)  \left( 1-\Lambda y_0 \right)^{2} y_0^{2}{\mu}^
{2}}} \right]
 \left( y-{  y_0} \right) ^{2} \nonumber \\
   && +a_{{3}}
 \left( y-{ y_0} \right) ^{3}+a_{{4}} \left( y- y_0 \right)
 ^{4}+\dots
\end{eqnarray}
where $a_0$, and $a_3$ are arbitrary constants, $a_4$ is determined
in terms of $a_0$ and $a_3$, and dots indicate higher order terms,
also completely determined in terms of $a_0$, and $a_3$. This result
is, in fact, more general for the type of equations considered here,
as is shown in Appendix A. \\

We have then a situation where Eq. (\ref{diag40}) has solutions that
are regular in $0 < y < 1/\Lambda$, and such that, for appropriate
values of $\Omega$, are also finite at the boundaries $y=0$, and
$y=1/\Lambda$, but we cannot ascertain whether the solutions form a
complete set, or if the eigenvalues $\Omega^2$ are, for instance,
bounded from below. At this point we recall that there is a well
known procedure that may allow us to establish a map between the
solutions of (\ref{diag40}) and those of a related self adjoint
problem. This is the Darboux transformation \cite{darboux}, also
considered as the introduction of an ``intertwining'' operator
\cite{price}, which will be used here in a manner similar to that
analyzed in \cite{dotti}. The explicit construction is shown in the
next Section.

\section{The Darboux transformation.}

Consider, in a given domain $r_0 \leq r \leq r_1$ of $r$,  the
solutions $\phi=\phi(r)$ of the equation,
\begin{equation}\label{sp02}
    -\frac{d^2\phi}{dr^2}+V(r)\phi = \Omega^2 \phi
\end{equation}
and a particular solution,
\begin{equation}\label{sp04}
    -\frac{d^2\phi_0}{dr^2}+V(r)\phi_0 = \Omega_0^2 \phi_0
\end{equation}

In correspondence with a given solution $\phi(r)$, (no particular
boundary conditions implied), define a new function $\chi(r)$, given
by,
\begin{equation}\label{sp06}
    \chi(r)= \frac{d \phi}{dr} -\frac{\phi}{\phi_0}\frac{d
    \phi_0}{dr}
\end{equation}

Then, a simple derivation shows that $\chi(r)$ is a solution of,
\begin{equation}\label{sp08}
     -\frac{d^2\chi}{dr^2}+\widetilde{V}(r)\chi = \Omega^2 \chi
\end{equation}
where,
\begin{equation}\label{sp10}
      \widetilde{V}(r) = -V(r)+2
      \Omega_0^2+\frac{2}{\phi_0^2}\left(\frac{d\phi_0}{dr}\right)^2
\end{equation}

Suppose now that $\widetilde{V}(r)$ is such that (\ref{sp08}) admits
a self adjoint extension. This implies that, after imposing
appropriate boundary conditions, there will be a complete set of
solutions of (\ref{sp08}) (``eigenfunctions'') $\chi_{\alpha}$, with
corresponding ``eigenvalues'' $\Omega_{\alpha}^2$, and satisfying an
orthonormality condition,
\begin{equation}\label{sp12}
      \int_{r_0}^{r_1}{\chi^*_{\alpha}\chi_{\beta} dr} =
      \delta_{\alpha,\beta}
\end{equation}
with $\delta_{\alpha,\beta}$ a Kronecker or Dirac delta function as
appropriate.

 Next, using (\ref{sp06}) we find,
\begin{equation}\label{sp18}
(\Omega_0^2-\Omega_{\alpha}^2)\phi_{\alpha}(r)= \frac{d
\chi_{\alpha}}{dr} +\frac{\chi_{\alpha}}{\phi_0}\frac{d\phi_0}{dr}
\end{equation}

This implies that corresponding to every $\chi_{\alpha}$, with
$\Omega_{\alpha}^2 \neq \Omega_0^2$, we will have a (possibly
singular) solution $\phi_{\alpha}$ of (\ref{sp02}).

Going back to (\ref{diag52}), (\ref{diag54}) of the previous
Section, identifying $\widetilde{W}(r)$ with $\phi(r)$, and using
(\ref{diag54}), we see that $\widetilde{V}(r)$, is given implicitly
in part by the coefficients $Q_i$, and in part by the chosen
solution $\phi_0(r) \equiv \widetilde{W}^{(0)}(r)$, where
$\widetilde{W}^{(0)}(r)$ is a solution of (\ref{diag52}) with
$\Omega^2=\Omega_0^2$. But, in view of (\ref{diag44}) and,
(\ref{diag48}), we have,
\begin{equation}\label{sp20}
   \widetilde{W}^{(0)}\left(r(y)\right) = \frac{W^{(0)}(y)}{K(y)}
\end{equation}
and,
\begin{eqnarray}
\label{sp22}
  \frac{1}{\phi_0}\frac{d\phi_0}{dr} & = &  \frac{K} {W^{(0)}}
  \frac{dy}{dr}\frac{d}{dy}\left( \frac {W^{(0)}}{K}\right)
  \nonumber \\
   &=&
   \frac{1}{\sqrt{Q_3}}\left(\frac{1}{W^{(0)}}
   \frac{d W^{(0)}}{dy}-\frac{1}{K}
   \frac{d K}{dy} \right) \\
   & = &\frac{1}{\sqrt{Q_3}}\left(\frac{1}{W^{(0)}}\frac{d
   W^{(0)}}{dy}-\frac{1}{2} Q_1+
   \frac{1}{4Q_3} \frac{d Q_3}{dy}\right) \nonumber
\end{eqnarray}
and, therefore, we have,
\begin{eqnarray}
\label{sp24}
  \widetilde{V}(r(y)) & = & -V(r(y))+2
      \Omega_0^2 \\
      & & +\frac{2}{Q_3}\left(\frac{1}{W^{(0)}}\frac{d
   W^{(0)}}{dy}-\frac{1}{2} Q_1+
   \frac{1}{4Q_3} \frac{d Q_3}{dy}\right)^2 \nonumber
\end{eqnarray}
where $V(r(y)) \equiv \mathbf{V}(r)$ is given by (\ref{diag54}). As
shown in more generality in Appendix A, and Appendix B, given the
form of the functions $Q_i$ near the singular point $y=y_s$, and
considering a general solution $W^{(0)}(y)$ that does not vanish at
$y=y_s$, the resulting $\widetilde{V}(r(y))$ is regular in the
neighbourhood of $y=y_s$. Since the functions $Q_i$ are regular in
$0 < y <y_s$, and in $y_s<y<1/\Lambda$, $\widetilde{V}(r(y))$ will
be regular in $0<y<1/\Lambda$ if $W^{(0)}(y)$ has no zeros in that
interval. Let us assume that such solution exists, and that the
resulting $\widetilde{V}(r)$ is such that (\ref{sp08}) admits a self
adjoint extension. Then, since the interval $(r(0),r(1/\Lambda))$ is
finite, the spectrum is entirely discrete, and we may take the
functions $\chi_{\alpha}$ as real, and normalized to a Kronecker
delta function.

We may now take for the solutions $W^{(\alpha)}$ the set,
\begin{eqnarray}
\label{sp26}
  W^{(\alpha)}(y) &=&  K(y)\widetilde{W}^{(\alpha)}(r(y))
  \nonumber  \\
   &=& K(y)\frac{1}{\Omega_0^2-\Omega_{\alpha}^2}\left(\frac{d
\chi_{\alpha}}{dr}
+\frac{\chi_{\alpha}}{\phi_0}\frac{d\phi_0}{dr}\right) \\
&=&K(y)\frac{1}{\Omega_0^2-\Omega_{\alpha}^2}\left(\frac{d
\chi_{\alpha}}{dr} +\chi_{\alpha}
\left(\frac{1}{W^{(0)}}\frac{dW^{(0)}}{dy}-\frac{1}
{K}\frac{dK}{dy}\right)
  \left(\frac{dr}{dy}\right)^{-1} \right) \nonumber
\end{eqnarray}
which will also be real, but we must impose the additional
requirement that the $ W^{(\alpha)}(y)$ satisfy the boundary
conditions (\ref{diag26})-(\ref{diag28}). Since from (\ref{sp06}) we
have,
\begin{equation}\label{sp28}
    \chi_{\alpha}(r(y))= \frac{1}{K W^{(0)}}\left( W^{(0)}
    \frac{dW^{(\alpha)}}{dy}- W^{(\alpha)} \frac{d
    W^{(0)}}{dy}\right) \left(\frac{dr}{dy}\right)^{-1},
\end{equation}
This implies, as can be checked, that if $ W^{(0)}(y)$ diverges as
$\ln(y)$, or $\ln(1-\Lambda y)$, at, respectively, $y=0$, or
$y=1/\Lambda$, then, $\chi(r)$ will diverge as $1/(\sqrt{r}
\ln(r))$, or $1/(\sqrt{(r_1-r)} \ln(r_1-r))$, which leads to
complications that are outside the present discussion. The simplest
choice, remember that we are interested in the properties of $W(y)$,
and not on those of the $\chi^{(\alpha)}$, is to impose that
$W^{(0)}(y)$ satisfies also the boundary conditions
(\ref{diag26})-(\ref{diag28}). This, as shown in Appendix C, leads
to a unique self adjoint extension for (\ref{sp08}). But notice that
in this case, $\Omega_0^2$ must be in the spectrum of
$W^{(\alpha)}(y)$, and, therefore, in accordance with (\ref{sp06}),
{\em not} in the spectrum of $\chi_{\alpha}$.

Let us consider now $\alpha$ and $\beta$ such that
$\Omega^2_{\alpha} \neq \Omega_0^2$, and $\Omega_{\beta}^2 \neq
\Omega_0^2$. We then have,

\begin{equation}\label{sp30}
    \int_0^1 \frac{1}{K^2}\left(
    \frac{dW^{(\alpha)}}{dy}- \frac{W^{(\alpha)} }{W^{(0)}}\frac{d
    W^{(0)}}{dy}\right)
    \left(
    \frac{dW^{(\beta)}}{dy}- \frac{W^{(\beta)}}{W^{(0)}} \frac{d
    W^{(0)}}{dy}\right)
    \left(\frac{dr}{dy}\right)^{-1} dy =
    \mathcal{N}_{\alpha}\delta_{\alpha,\beta}
\end{equation}
where,
\begin{equation}\label{sp30a}
   \mathcal{N}_{\alpha}= \int_0^1 \frac{1}{K^2}\left(
    \frac{dW^{(\alpha)}}{dy}- \frac{W^{(\alpha)} }{W^{(0)}}\frac{d
    W^{(0)}}{dy}\right)^2
    \left(\frac{dr}{dy}\right)^{-1} dy
\end{equation}

This will be applied in the next Section to obtain the general
evolution of an arbitrary perturbation.

\section{General time evolution of the perturbations.}

Consider again (\ref{diag32}). We define the linear
(integro-differential) operator $\mathcal{O}$ as,
\begin{equation}
\label{gte01}
    \mathcal{O}\left(W\right)\equiv \int_0^y{\left[{\frac {
 \left( \kappa+\mu-1 \right) {W}}{2 \mu F^{g}_4 {\cal{K}}
   \left( F^{g}_3 + F^{g}_4
   \right) y' \left( 1- \Lambda y' \right) }}+
  \frac{1}{(F^{g}_3 + F^{g}_4)}
   {\frac {d}{dy'}} \left( {\frac {{W} }{F^{g}_4 {\cal{K}}  }}
    \right)\right] dy'}
\end{equation}

We now go back to the expansion (\ref{gau03a}), but restricted to
$\ell=0$, and a particular value of $k$. We take as the allowed
values of $\Omega$ the full set of $\Omega_{\alpha}$ such that
$\Omega_{\alpha}^2$ are the eigenvalues of the self adjoint problem
associated to $W$, plus $\Omega_0$ such that $\Omega_0^2$ is the
eigenvalue corresponding to $W^{(0)}$, which is not included in the
set $\{\Omega_{\alpha}\}$. The general expansion (\ref{gau03a}), in
the case of $F_4$, then takes the form,
\begin{eqnarray}\label{grt01a}
     h_{\phi \phi} & = &\sum_{\alpha}
      {\left[\exp\left(i \Omega_{\alpha} t \right)
    \mathcal{A}_{\alpha}+
    \exp\left(-i \Omega_{\alpha} t \right)
    \mathcal{B}_{\alpha}\right] F_4^{(\alpha)}\left(y\right)} \\
    & & +\left[\exp\left(i \Omega_{0} t \right)
    \mathcal{A}_{0}+
    \exp\left(-i \Omega_{0} t \right)
    \mathcal{B}_{0}\right] F_4^{(0)}\left(y\right) \nonumber,
\end{eqnarray}
In these expressions $\Omega_{{\alpha}}$ ($\Omega_0$) is either the
positive real root or positive imaginary root of $\Omega_{{
\alpha}}^2$, ($\Omega_{0}^2$).  $\mathcal{A}_{{\alpha}}$, and
$\mathcal{B}_{{\alpha}}$ are arbitrary constants, and,
\begin{equation}
\label{grt01b}
    F_4^{(a)}\left(y\right)= F^{g}_4(y)
    \mathcal{O}\left(W^{a}\right) +C_{a} F^{g}_4
\end{equation}
where $a$ stands for either $\alpha$, or $0$, with $C_{a}$ an
arbitrary constant.

We can see that these expressions are not very useful, because the
function $F_4$, (and in general all $F_i$), contains gauge dependent
parts, and these imply that $h_{\phi \phi}$, (and in general
$h_{\mu\nu}$), will contain time dependencies that are not fully
determined by the initial conditions, in correspondence with the
gauge ambiguities present at any given time. Any physically relevant
information is therefore contained in the gauge invariant part of
the $h_{\mu\nu}$, and this, in principle, can be extracted precisely
from, e.g., $W$. More explicitly, consider again (\ref{grt01a}). In
view of the linearity of $\mathcal{O}$, and the fact that
$F^{g}_4(y)$ does not depend on $\Omega$, this can be written as,
\begin{equation}
\label{grt01c}
     h_{\phi \phi}(t,y)= F^{g}_4(y)
     \mathcal{O}\left(\mathbf{W}(t,y)\right)
     +F^{g}_4(y)\mathbf{C}(t,y)
\end{equation}
where,
\begin{equation}
\label{grt01d}
 \mathbf{C}(t,y)= \sum_{a} {C_{a}
 \left[\exp\left(i \Omega_{a} t \right)
    \mathcal{A}_{a}+
    \exp\left(-i \Omega_{a} t \right)
    \mathcal{B}_{a}\right] },
\end{equation}
and,
\begin{equation}\label{sp32}
    \mathbf{W}\left(t,y\right)= \sum_{a}
    {\left[\exp\left(i \Omega_{a} t \right)
    \mathcal{A}_{a}+
    \exp\left(-i \Omega_{a} t \right)
    \mathcal{B}_{a}\right] W^{(a)}\left(y\right)},
\end{equation}
where the index $a$ extends over all the $\alpha$, plus $0$.

Using now (\ref{diag40}) we have,
\begin{eqnarray}
\label{gte04}
  \frac{\partial^2 \mathbf{W}}{\partial t^2} &=& \sum_{a}
   {\left[\exp\left(i \Omega_{a} t \right)
    \mathcal{A}_{a}+
    \exp\left(-i \Omega_{a} t \right)
    \mathcal{B}_{a}\right]\left(-\Omega_{a}{}^2
     W^{(\alpha)}\right)} \nonumber \\
   &=&   \frac{1}{Q_3}\frac{\partial^2 \mathbf{W}}
   {\partial y^2}
     -\frac{Q_1}{Q_3}\frac{\partial \mathbf{W}}
     {\partial y} -\frac{Q_2}{Q_3} \mathbf{W}
\end{eqnarray}
which represents the general evolution equation for the gauge
invariant $\mathbf{W}$. Suppose we are given initial data for
(\ref{gte04}) in the form of the functions $\mathbf{W}(0,y)$, and
$(\partial \mathbf{W}(t,y)/\partial t)|_{t=0}$. Using
(\ref{sp30},\ref{sp30a}), for $\alpha$ in the set of eigenvalues of
$\chi_{\alpha}$, we easily obtain,
\begin{eqnarray}
\label{sp34a}
  \mathcal{A}_{\alpha}+\mathcal{B}_{\alpha} &=&
  \frac{1}{\mathcal{N}_{\alpha}}
   \int_0^1\frac{1}{K^2}\left(
    \frac{dW^{(\alpha)}}{dy}- \frac{W^{(\alpha)} }{W^{(0)}}\frac{d
    W^{(0)}}{dy}\right)  \\
    & & \times \left.\left(
    \frac{\partial \mathbf{W}}{\partial y}-
     \frac{\mathbf{W} }{W^{(0)}}\frac{d
    W^{(0)}}{dy}\right)\right|_{t=0}
    \left(\frac{dr}{dy}\right)^{-1} dy \nonumber
\end{eqnarray}
and,
\begin{eqnarray}
\label{sp34b}
  \mathcal{A}_{\alpha}-\mathcal{B}_{\alpha} &=&
  \frac{1}{i \Omega_{\alpha}\mathcal{N}_{\alpha}}
   \int_0^1\frac{1}{K^2}\left(
    \frac{dW^{(\alpha)}}{dy}- \frac{W^{(\alpha)} }{W^{(0)}}\frac{d
    W^{(0)}}{dy}\right)  \\
    & & \times \left.\left(
    \frac{\partial^2 \mathbf{W}}{\partial t \partial y}-
    \frac{\partial \mathbf{W} }{\partial t} \frac{1}{W^{(0)}}\frac{d
    W^{(0)}}{dy}\right)\right|_{t=0}
    \left(\frac{dr}{dy}\right)^{-1} dy \nonumber
\end{eqnarray}
from which we can straightforwardly solve for
$\mathcal{A}_{\alpha}$, and $\mathcal{B}_{\alpha}$. Finally we may
obtain the remaining coefficients $\mathcal{A}_{0}$, and
$\mathcal{B}_{0}$, using
\begin{equation}\label{sp36a}
  \left(\mathcal{A}_{0}+\mathcal{B}_{0}\right) W^{(0)}(y)= \mathbf{W}(0,y)-
  \sum_{\alpha}{\left(\mathcal{A}_{\alpha}+\mathcal{B}_{\alpha}\right)
  W^{(\alpha)}(y)}
\end{equation}
and,
\begin{equation}\label{sp36b}
  \left(\mathcal{A}_{0}-\mathcal{B}_{0}\right) W^{(0)}(y)= \frac{1}{i
  \Omega_0}\left[\left.\frac{\partial \mathbf{W}}{\partial t}\right|_{t=0}-
  \sum_{\alpha}{i\Omega_{\alpha}\left(\mathcal{A}_{\alpha}-\mathcal{B}_{\alpha}\right)
  W^{(\alpha)}(y)}\right]
\end{equation}
Therefore, the expansion (\ref{sp32}) is complete, in the sense that
it provides the time of evolution of an arbitrary linear
perturbation in terms of the ``modes'', that is, the solutions of
(\ref{diag40}) that satisfy the appropriate boundary conditions.
Thus, if one, or more, of these modes correspond to eigenvalues
$\Omega^2 <0$, we must conclude that the system is {\em generically
unstable.}

In the next Section we will provide numerical evidence for the
existence of such unstable modes for the system
(\ref{diag04a},\ref{diag04b},\ref{diag06}).

\section{Numerical results.}

In this Section we consider numerical solutions of the perturbation
equations. Because of their singular nature we cannot directly
impose conditions at the boundaries at $y=0$ or $y=1/\Lambda$.
Instead, we consider, where possible, appropriate expansions
corresponding to the chosen type of solution, to impose initial data
near the corresponding boundary, and proceed to a numerical
integration using a Runge - Kutta integration method, after fixing
the values of $\Lambda$, $\kappa$, and $k$, to explore a range of
values of $\Omega$, looking for solutions that satisfy the desired
boundary conditions at both $y=0$ or $y=1/\Lambda$.

More explicitly, let us consider $\kappa=1/2$. In this case the
perturbation equations reduce to,
\begin{eqnarray}
\label{num02}
  \frac{dF_1}{dy} &=& -\frac{dF_4}{dy}-\frac{5 F_1}{14 y (1-\Lambda y)}
  +\frac{(28 \Lambda y -51) F_4}{42 y (1-\Lambda y)}, \\
  \frac{dF_3}{dy} &=& -\frac{dF_4}{dy}-\frac{5 F_3}{14 y (1-\Lambda y)}
  +\frac{(4 \Lambda y -3) F_4}{6 y (1-\Lambda y)}
  \nonumber
\end{eqnarray}
and,
\begin{eqnarray}
\label{num04}
 {\frac {d F_4}{dy}}&=& \frac{7  \left( F_{{3}}  +F_{{4}}
  \right) {\Omega}^{2}}{2 \left( 1-
\Lambda\,y \right) ^{{\frac {5}{21}}}{y}^{3/7} \left(9 -
7\,\Lambda\,y \right)}
 - \frac{7 {y}^{2/7} \left( F_{{1}}  +F_{{4} }  \right)
  {k}^{2}}{2 \left( 1-\Lambda\,y
\right) ^{{ \frac {20}{21}}} \left( 9-7\,\Lambda\,y \right)}  \\
 & &+
{\frac { \left(180 -420\,\Lambda y \right) F_{{1}}  +
\left( 420
\Lambda y-405 \right) F_{{3}} +
 \left( 392 {\Lambda}^{2}{y}^{2}-1932
\Lambda y+1179
 \right) F_{{4}} }{168 y \left( 1-\Lambda\,y \right)
 \left( 7\,\Lambda\,y-9 \right) }} \nonumber
\end{eqnarray}

 We also have,
\begin{equation}\label{num12}
    W(y)=  \frac {45-35 \Lambda y}{8} F_{{3}}
 \left( y \right) +  \frac{15+35 \Lambda y -98
 (1-\Lambda y)^{\frac{1}{21}}}{8} F_{{4}} \left( y \right)
\end{equation}
and we notice that we can derive an expression for $dW/dy$ entirely
in terms of the $F_i$, (with no derivatives of the $F_i$), using
(\ref{num12}), and the perturbation equations
(\ref{num02})-(\ref{num04}), although we shall not show it here for
simplicity. It is straightforward to derive now an expression for
$\widetilde{V}(y)$, entirely in terms of the $Q_i$, and the $F_i$
again with no derivatives of the $F_i$, although the resulting
expression is rather too lengthy to be displayed here. It was,
nevertheless, used to compute $\widetilde{V}(y)$ numerically, using
the results of the numerical integration of the $F_i$.

To solve the problem of finding numerical solutions of the
perturbation equations satisfying appropriate boundary conditions we
used a ''shooting'' approach, imposing finite boundary conditions at
$y=0$, and looking for solutions that satisfy the required
conditions as we approach $y=1/\Lambda$.  A straightforward
computation shows that near $y=0$, for general $k$, $\Lambda$, and
$\Omega$, we have a solution that admits an expansion of the form,
\begin{eqnarray}
\label{num06}
 F_{{1}} \left( y \right) & = & a_{{0}} \left( 1+{\frac
{175}{816}} {\Omega }^{2}{y}^{4/7}-{\frac {50}{459}} \Lambda
y-{\frac {539539}{4543488}}  {\Omega}^{4}{y}^{{\frac {8}{7}}}+{\frac
{22295}{128061}} {k}^{2}{y} ^{{\frac {9}{7}}} \right. \nonumber
\\ & &   +{\frac
{424621}{3998808}} \Lambda {\Omega}^{2}{y}^ {{\frac {11}{7}}}
+{\frac {1191431423}{72623112192}} {\Omega}^{6}{y}^{ {\frac
{12}{7}}}-{\frac {516215}{1038830832}} {k}^{2}{\Omega}^{2}{y}^
{{\frac {13}{7}}} \\ & & \left. -{\frac {11200}{169371}}
{\Lambda}^{2}{y}^{2} +\dots
 \right), \nonumber
\end{eqnarray}

\begin{eqnarray}
\label{num08}
 F_{{3}} \left( y \right) & = & a_{{0}} \left( -{\frac {7}{17}}
 +{\frac {455}
{816}} {\Omega}^{2}{y}^{4/7}-{\frac {350}{1377}} \Lambda y-{\frac {
717899}{4543488}} {\Omega}^{4}{y}^{{\frac {8}{7}}}+{\frac
{28175}{ 128061}} {k}^{2}{y}^{{\frac {9}{7}}} \right. \nonumber \\
 & & +{\frac {4667}{16456}} \Lambda {\Omega}^{2}{y}^{{\frac
{11}{7}}}+{\frac {47412547}{2504245248}} { \Omega}^{6}{y}^{{\frac
{12}{7}}}-{\frac {18865}{33510672}} {k}^{2}{ \Omega}^{2}{y}^{{\frac
{13}{7}}} \\ & & \left.-{\frac {30800}{169371}}\,{\Lambda}^{2
}{y}^{2} \dots  \right) \nonumber
\end{eqnarray}
and,

\begin{eqnarray}
\label{num10}
 F_{{4}} \left( y \right) & = & a_{{0}} \left( -{\frac {5}{17}
 }-{\frac {91}{
816}} {\Omega}^{2}{y}^{4/7}-{\frac {10}{459}} \Lambda y+{\frac {
343343}{4543488}} {\Omega}^{4}{y}^{{\frac {8}{7}}}-{\frac {14651}{
128061}} {k}^{2}{y}^{{\frac {9}{7}}} \right. \nonumber \\
 & &-{\frac
{689}{8712}}\,\Lambda\,{ \Omega}^{2}{y}^{{\frac {11}{7}}}-{\frac
{29059303}{2504245248}} { \Omega}^{6}{y}^{{\frac {12}{7}}}+{\frac
{12005}{33510672}} {k}^{2}{ \Omega}^{2}{y}^{{\frac {13}{7}}}
\\
 & & \left.+{\frac {560}{169371}}\,{\Lambda}^{2}{ y}^{2}\dots  \right)
\nonumber
\end{eqnarray}
where $a_0$ is a constant and dots indicate higher order terms. This
expansion corresponds to the solutions where the $F_i(y)$ approach a
finite limit as $y \to 0$.\\

Using the expansions (\ref{num06})-(\ref{num08})-(\ref{num10}), to
fix initial values near $y=0$, we carried out a Runge - Kutta
integration of the system (\ref{num02})-(\ref{num04}), paying
attention to the behaviour of $W(y)$, as $y \to 1/\Lambda$. In
accordance with the ``shooting'' idea, for fixed $k$ (and setting
$\Lambda=1$, as already indicated), we looked for values of
$\Omega^2$ such that $W(y)$ approached a finite value as $y \to
1/\Lambda$, but showed a divergent behaviour as we changed
$\Omega^2$ to slightly larger or smaller values. Setting
$\kappa=1/2$, $\Lambda=1$, and $k=1$, we found that for the smaller
of those values, which was $\Omega^2=-0.6107$, ($\Omega = 0.7815
i$), we had that $W(y)$ did not vanish in $0 < y < 1/\Lambda$. We,
therefore, identified that solution with $W_0(y)$ of the previous
Section, and computed $\widetilde{V}(y(r))$. A plot of $W_0(y)$ as a
function of $y$, and of $\widetilde{V}$, as a function of $r$ (given
by (\ref{diag56}), with $\kappa=1/2$), both for $\Lambda=1$, $k=1$,
and $\Omega=0.7815 i$, are given in Fig.1, and Fig.2. Notice that
$\widetilde{V}(r)$ is bounded from below, in fact is positive
definite, and diverges to $+\infty$, both at $r=0$, and $r=r_1$.
Therefore, as discussed in the previous Section, the spectrum
$\Omega^2_{\alpha}$ is discrete and positive definite, but the
complete spectrum of $W(y)$ contains also the imaginary eigenvalue
$\Omega_0 = 0.7815 i$. These results imply that for $\kappa=1/2$,
(and, from (\ref{intro12}), also for $\kappa=1/4$), the linear
evolution of a general perturbation will contain an unstable mode,
and, therefore, the space time will be linearly unstable.

\begin{figure}[H]
\centerline{\includegraphics[height=12cm,angle=-90]{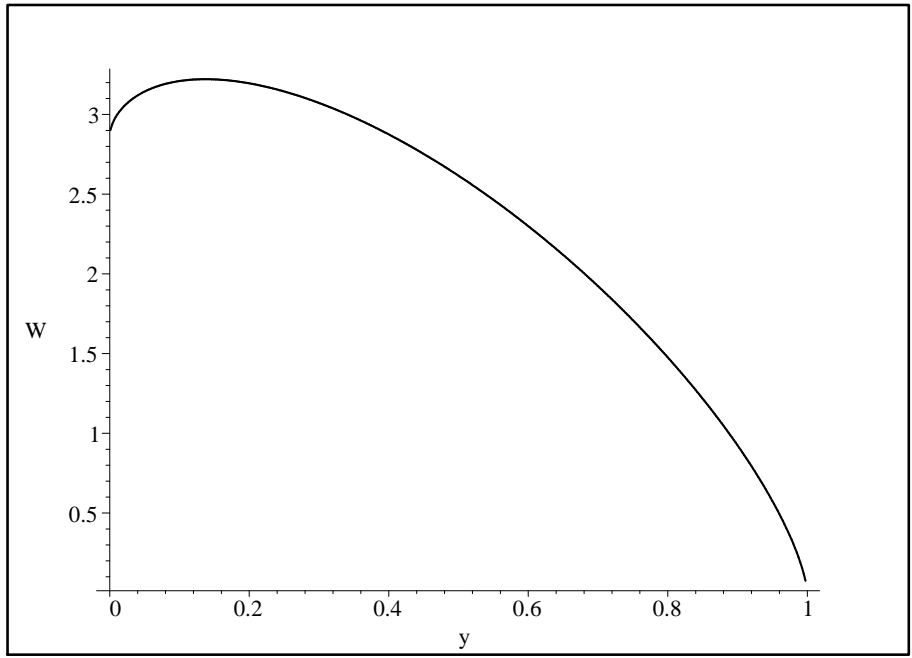}}
\vspace{0cm} \caption{A plot of $W(y)$, computed using the numerical
integration of the $F_i(y)$, as functions of $y$, for $\kappa=1/2$,
$k=1$, $\Lambda=1$ and $\Omega^2=-0.6107$. This is identified with
$W_0$, and used to compute $\widetilde{V}$.}
\end{figure}

\begin{figure}[H]
\centerline{\includegraphics[height=12cm,angle=-90]{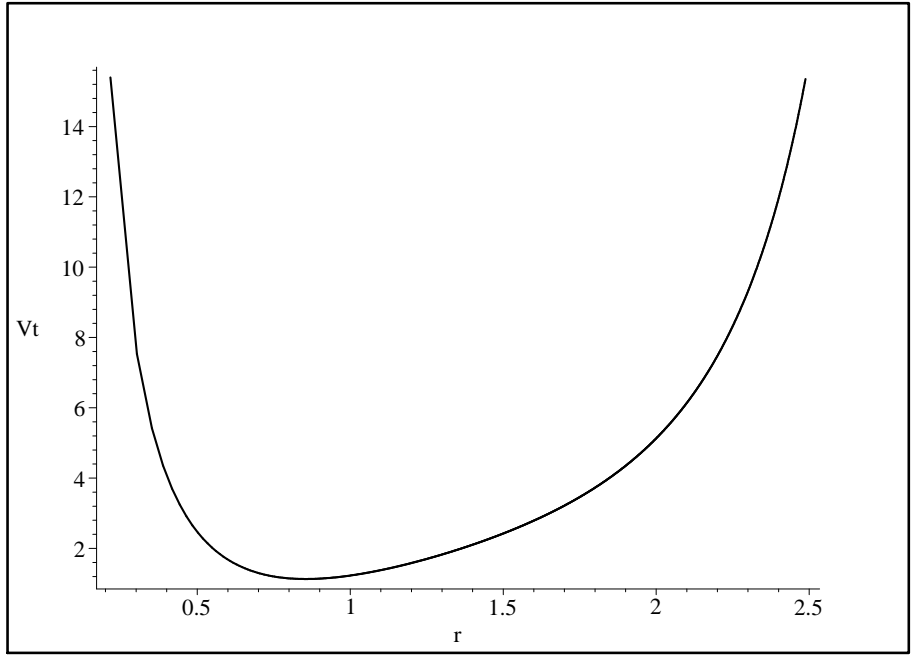}}
\vspace{0cm} \caption{A plot of $\widetilde{V}(r)$, as a function of
$r$, computed using the numerical integration of the $F_i(y)$, as
functions of $y$, for $\kappa=1/2$, $k=1$, $\Lambda=1$ and
$\Omega^2=-0.6107$, identifying $W_0(y)$ with $W(y)$ computed for
the same values of the parameters.}
\end{figure}

We also computed, as a check, the next higher eigenfunction, which
is plotted in Fig.3. obtaining $\Omega^2=3.853$, ($\Omega= 1.963$).
This is in qualitative correspondence to the lowest eigenvalue for
$\widetilde{V}$, as shown in Fig. 2., which we would expect to be
around $\Omega^2 \sim 4$.

\begin{figure}[H]
\centerline{\includegraphics[height=12cm,angle=-90]{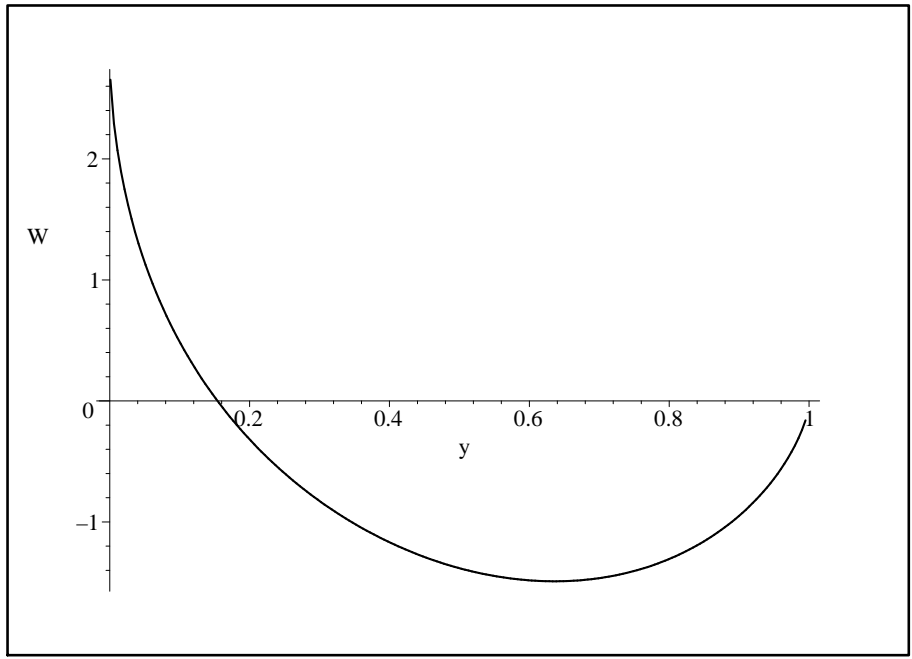}}
\vspace{0cm} \caption{A plot of $W(y)$, computed using the numerical
integration of the $F_i(y)$, as functions of $y$, for $\kappa=1/2$,
$k=1$, $\Lambda=1$ and $\Omega^2=3.853$.}
\end{figure}

We have also carried out the integration, again for $\Lambda =1 $,
and $k=1$, and entirely along the same lines, for $\kappa=1/3$,
(isometric to $\kappa=2/5$), obtaining $\Omega^2= -0.6545$,
($\Omega=0.809 i$), for the lowest eigenvalue ($\Omega_0$), and
$\Omega^2= 5.128$, for the next eigenvalue, with qualitatively
similar graphs for $W$, and $\widetilde{V}$ as those for
$\kappa=1/2$. Similar results were obtained for $\kappa=3/4$,
(isometric to $\kappa = 1/7$), with $\Omega^2=-0.7075$,
($\Omega=0.8411 i$) for the lowest eigenvalue $\Omega_0$, and
$\Omega^2=2.091$ for the next higher eigenvalue, again with graphs
for $W$, and $\widetilde{V}$ similar to those for $\kappa=1/2$. All
these results, therefore confirm the linear instability of the Linet
- Tian metrics with a positive cosmological constant.

In the next Section we consider some special values of $\kappa$,
where a separate analysis is required.

\section{Some special cases.}

In this Section we consider two particular values of $\kappa$. They
are $\kappa=0$, and $\kappa=1$. The reason why these values are
particular can be seen from (\ref{diag10}, \ref{diag12}), and
(\ref{diag16}, \ref{diag16a}), which show that there is a
qualitative change in the behaviour of the $F_i$, in the limits $y
\to 0$, and $y \to 1/\Lambda $, and, therefore, in that of $W(y)$.
In what follows we consider these cases separately.

\subsection{The case $\kappa=0$.}

If we take $\kappa=0$ in the perturbation equations
(\ref{diag04a},\ref{diag04b},\ref{diag06}), we get,
\begin{equation}\label{ka001}
    \frac{dF_1}{dy}= -\frac{dF_4}{dy}-\frac{3-2\Lambda y}{3 y
    (1-\Lambda y)}F_4,
\end{equation}
\begin{equation}\label{ka002}
    \frac{dF_3}{dy}= -\frac{dF_4}{dy}-\frac{3-2\Lambda y}{3 y
    (1-\Lambda y)}F_4,
\end{equation}
and,
\begin{equation}\label{ka003}
    \frac{d^2F_4}{dy^2}= \frac{2 \Lambda y-1}{y(1-\Lambda y)}
    \frac{dF_4}{dy}
    -\frac{\Omega^2 F_4}{3 y (1-\Lambda y)^{5/3}} +\frac{9-6\Lambda
    y -2 \Lambda^2 y^2 +3 k^2 y (1-\Lambda y)^{1/3}}{9 y^2
    (1-\Lambda y )^2} F_4
\end{equation}

This implies first that there is no gauge ambiguity other than an
additive constant in $F_1$, and $F_3$. We also notice that once
(\ref{ka003}) is solved for $F_4(y)$, the other equations are solved
by direct integration. Regarding (\ref{ka003}) itself, the
coefficients are regular in $ 0 < y < 1/\Lambda$, and diverge both
at $y=0$, and $y=1/\Lambda$. We may use the standard procedure to
put it in self adjoint (Schroedinger like) form, by changing to a
new function $\widetilde{F}(r)$, such that,
\begin{equation}\label{ka005}
    F_4(y)= K_0(y) \widetilde{F}(r(y))
\end{equation}
where,
\begin{equation}\label{ka007}
    K_0(y) =\frac{1}{y^{1/4} (1-\Lambda y)^{1/12}}
\end{equation}
and,
\begin{equation}\label{ka009}
    \frac{dr}{dy} =\frac{1}{\sqrt{3} y^{1/2} (1-\Lambda y)^{5/6}}
\end{equation}

Replacing in (\ref{ka003}), we finally obtain,
\begin{equation}\label{ka011}
    -{\frac {d^{2} \widetilde{F}}{d{y}^{2}}}
 + \widetilde{V}\widetilde{F} = \Omega^{2}
\widetilde{F}
\end{equation}
where $\widetilde{V}(r)$ is given implicitly by,
\begin{equation}\label{ka013}
   \widetilde{V}(r)=
 {\frac { 16 {k}^{2} y (1-\Lambda y)^{1/3}+45 -40\Lambda y }{16 y \left(
1-\Lambda y \right) ^{1/3}}}
\end{equation}
We can see that the ``potential" $\widetilde{V}(r)$ is positive
definite and continuous in $0 < y < 1/\Lambda$, and that it diverges
to $+\infty$, for both $y \to 0$, and $y \to 1/\Lambda$. It is
straightforward to prove that (\ref{ka011}) admits a unique self
adjoint extension, and, on account of the properties of
$\widetilde{V}$, with a positive definite spectrum. This does not
immediately imply that there are no unstable modes for $\kappa=0$,
because we have restricted to $\ell=0$. In the next Subsection we
analyze the case $\kappa=1$, and find unstable modes. On account of
the symmetries between metrics with $\kappa=0$, and with $\kappa=1$,
this implies that there are also unstable modes for $\kappa=0$, but
for perturbations with $\ell \neq 0$.

\subsection{The case $\kappa=1$.}

In the case $\kappa=1$, the differential equation for $W(y)$ takes
the form,
\begin{eqnarray}
\label{kap101} -{\frac {d^{2}W}{d{y}^{2}}} +{\frac {
 \left( 4{y}^{4/3}{k}^{2} \left( 2\Lambda y-5 \right) +6\Lambda
y+3 \right)   }{ 3\left( 4{y }^{4/3}{k}^{2}-4\Lambda y+1
\right) y \left( 1-\Lambda y \right) }}{\frac {d W}{dy}} & & \\
 +{\frac {
\left( 4{k}^{4}{y}^{5/3}-{y}^{1/3} \left( -20\, \Lambda
y+15+8\,{y}^{2}{\Lambda}^{2} \right) {k}^{2}+6\Lambda
 \left( 1-\Lambda y \right)  \right)   }{3
 \left( 1-\Lambda y \right) ^{2}y \left( 4{y}^{4/3}{k}^{2}-4
\Lambda y+1 \right) }}W & = & {\frac {{\Omega}^{2}}{3 \left(
1-\Lambda y \right) {y}^{5/3}}} W \nonumber
\end{eqnarray}

We first notice that if we define a new function $W_1$, such that
$W(y)=W_1(z(y))$, where $z(y)=\Lambda y$, then (\ref{kap101}) takes
the form,
\begin{eqnarray}
\label{kap102}
-{\frac {d^{2}W_1}{d{z}^{2}}} +{\frac {
 \left( 4{z}^{4/3}{k_1}^{2} \left( 2 z-5 \right) +6
z+3 \right)   }{ 3\left( 4{z }^{4/3}{k_1}^{2}-4 z+1
\right) z \left( 1- z \right) }}{\frac {d W_1}{dz}} & & \\
 +{\frac {
\left( 4{k_1}^{4}{z}^{5/3}-{z}^{1/3} \left( -20 z+15+8\,{z}^{2}
\right) {k_1}^{2}+6
 \left( 1- z \right)  \right)   }{3
 \left( 1- z \right) ^{2}z \left( 4{z}^{4/3}{k_1}^{2}-4
 z+1 \right) }}W_1 & = & {\frac {{\Omega_1}^{2}}{3 \left( 1- z
\right) {z}^{5/3}}} W_1 \nonumber
\end{eqnarray}
where $k_1=k/\Lambda^{2/3}$, and $\Omega_1=\Omega/\Lambda^{1/6}$.
Therefore, since we are interested in the existence of solutions
with $\Omega^2 < 0$, without loss of generality, we will set
$\Lambda =1$ in (\ref{kap101}) in what follows. Next, we notice that
for $k^2 > 3 /4$, the coefficients in (\ref{kap101}) are regular in
$0<y<1$. Then, for $k^2 > 3/4$, with the transformation of Section
V, we find for equation (\ref{diag54}) the potential,
\begin{eqnarray}
\label{kap103}
  \mathbf{V}(y) &=& \left[768 {y}^{4}{k}^{6}+48{y}^{8/3} \left( 29-56 y
 \right) {k}^{4}+24 {y}^{4/3} \left( 68 y-73+32 {y}^{2} \right) {k}
^{2} \right. \nonumber \\ && \left.+512{y}^{4}-896{y}^{3}-1-528
{y}^{2}+832y\right]   \\
&& \times \left[48{y}^{1/3}
 \left( 4\,{y}^{4/3}{k}^{2}-4\,y+1 \right) ^{2} \left(1-y \right)
 \right]^{-1} \nonumber
\end{eqnarray}
which is regular in $0<y<1$. In particular, near $y=0$ we have,
\begin{equation}\label{kap104}
    \mathbf{V}(y) = -\frac{1}{48 y^{1/3}} +
    \mathcal{O}\left(y^{2/3}\right)
\end{equation}
and, from (\ref{diag56}), for $\kappa=1$ this implies that near
$r=0$ we have,
\begin{equation}\label{kap105}
    \mathbf{V}(r) = -\frac{1}{4 r^2} +
    \mathcal{O}\left(r^4\right)
\end{equation}

Similarly, near $y=1$ we have,
\begin{equation}\label{kap106}
    \mathbf{V}(y) = \frac{16k^2-3}{16 (1-y)} +
    \mathcal{O}\left((1-y)^0\right)
\end{equation}
and, therefore, near $r=r_1$, in terms of $r$ we have,
\begin{equation}\label{kap107}
    \mathbf{V}(r) = \frac{16k^2-3}{12 (r_1-r)^2} +
    \mathcal{O}\left((r_1-r)^0\right)
\end{equation}

The form (\ref{kap105}) of $\mathbf{V}(r)$ implies that near $r=0$
the general solution $\widetilde{W}(r)$ of (\ref{diag52}) takes the
form,
\begin{equation}\label{kap108}
    \widetilde{W}(r) \simeq C_1 \sqrt{r} +C_2 \ln(r)\sqrt{r} +\dots
\end{equation}
where $C_1$, and $C_2$ are arbitrary constants. Similarly, near
$r=r_1$ we have, in general,
\begin{equation}\label{kap109}
    \widetilde{W}(r) \simeq C_3 (r_1-r)^{\frac{1}{2}+\frac{2\sqrt{3} k}{3}}
     +C_4 (r_1-r)^{\frac{1}{2}-\frac{2\sqrt{3} k}{3}} +\dots
\end{equation}
with $C_3$, and $C_4$ arbitrary constants. As far as obtaining a
self adjoint extension for (\ref{diag52}), the behaviour of
$\widetilde{W}$ at the boundary $r=0$ corresponds to the {\em limit
circle} case. As discussed in \cite{gibbons}, (see also
\cite{gleidott}), in this case we impose on the solutions the
condition $C_2=0$ at the boundary $r=0$. For the boundary $r=r_1$ we
obtain a self adjoint extension only if we impose the condition
$C_4=0$. These boundary conditions then provide a self adjoint
extension for (\ref{diag52}). Since the interval $0\leq r \leq r_1$
is finite, the spectrum is discrete, and, as can be shown, with the
condition $C_2=0$, it is also bounded from below.

Since with the given boundary conditions the problem (\ref{diag52})
is self adjoint, with an appropriate normalization, the set of
eigenfunctions $\widetilde{W}_{\alpha}(r)$ corresponding to the
eigenvalues $\Omega_{\alpha}$ is complete and the
$\widetilde{W}_{\alpha}(r)$  satisfy the orthonormality conditions,
\begin{equation}\label{kap110}
    \int_0^{r_1}\widetilde{W}_{\alpha}(r)\widetilde{W}_{\beta}(r) dr
    = \delta_{\alpha \beta}
\end{equation}

Recalling that we have $W_{\alpha}(y) =
K(y)\widetilde{W}_{\alpha}(r(y))$, this can be written as,
\begin{equation}\label{kap111}
    \int_0^{1}\frac{1}{K(y)^2} W_{\alpha}(y)W_{\beta}(y) \sqrt{Q_3}
    dy
    =  \sqrt{\mathcal{N}_{\alpha}\mathcal{N}_{\beta}}\delta_{\alpha \beta}
\end{equation}
where,
\begin{equation}\label{kap112}
  \mathcal{N}_{\alpha}=  \int_0^{1}\frac{1}{K(y)^2}
  \left(W_{\alpha}(y)\right)^2 \sqrt{Q_3} dy
\end{equation}

Therefore, instead of (\ref{diag52}) we may solve directly the
equation for $W_{\alpha}(y)$, without having to change to $r$. In
this case, the boundary conditions corresponding to the self adjoint
extension for $\widetilde{W}_{\alpha}(r)$ imply that for $y \to 0$,
$W_{\alpha}(y)$ admits an expansion of the form,
\begin{equation}\label{kap113}
  W_{\alpha}(y)= a_0\left[1-3\Omega^2 y^{1/3}+\frac{9}{4}\Omega^4
  y^{2/3}+\dots\right]
\end{equation}
where $a_0$ is a constant, while for $y \to 1$, the expansion takes
the form,
\begin{equation}\label{kap114}
  W_{\alpha}(y)= b_0 (1-y)^{\frac{k}{\sqrt{3}}}\left[1
  -\frac{6+3\Omega^2+5 \sqrt{3} k -2 k^2}{3(2\sqrt{3}k+3)}(1-y)
  +\dots\right]
\end{equation}
where $b_0$ is a constant. These expansions can be extended to
arbitrary orders in either $y$, or $1-y$. We have used
(\ref{kap113}), extended to order $y^2$, and (\ref{kap114}),
extended to $(1-y)^{\frac{k}{\sqrt{3}}+3}$, to solve (\ref{kap101})
numerically, (with $\Lambda=1$), for different values of $\Omega$,
and $k$, to obtain information on the spectrum of $\Omega$, for a
given value of $k$. In more detail, let us call $W_a(y,k,\Omega)$
the numerical solution of (\ref{kap101}), obtained by imposing
(\ref{kap113}), for the given values of $k$, and $\Omega$.
Similarly, $W_b(y,k,\Omega)$ is the numerical solution of
(\ref{kap101}) for the same $k$, and $\Omega$, but imposing
(\ref{kap114}). We also (arbitrarily) normalize these solutions so
that $W_a(0.5,k,\Omega)=W_b(0.5,k,\Omega)=1$. These solutions are
independent in general, but, for fixed $k$, we change the value of
$\Omega$ until we find that the plots of $W_a$, and $W_b$ as
functions of $y$ superimpose as perfectly as the graphic depiction
allows. The results, for two particular cases indicated in the
captions, are shown in Figures 4 and 5, where the curves plotted are
in fact the superposition of the (independent) solutions $W_a$, and
$W_b$, for the stated values of $k$, and $\Omega$. The reason for
using this procedure is that, because of the singular nature of the
boundaries, a simple ``shooting'' from either $y=0$ or $y=1$, trying
to impose the appropriate boundary condition at the opposite end,
turns out to be unreliable, and difficult to implement.

\begin{figure}[H]
\centerline{\includegraphics[height=12cm,angle=-90]{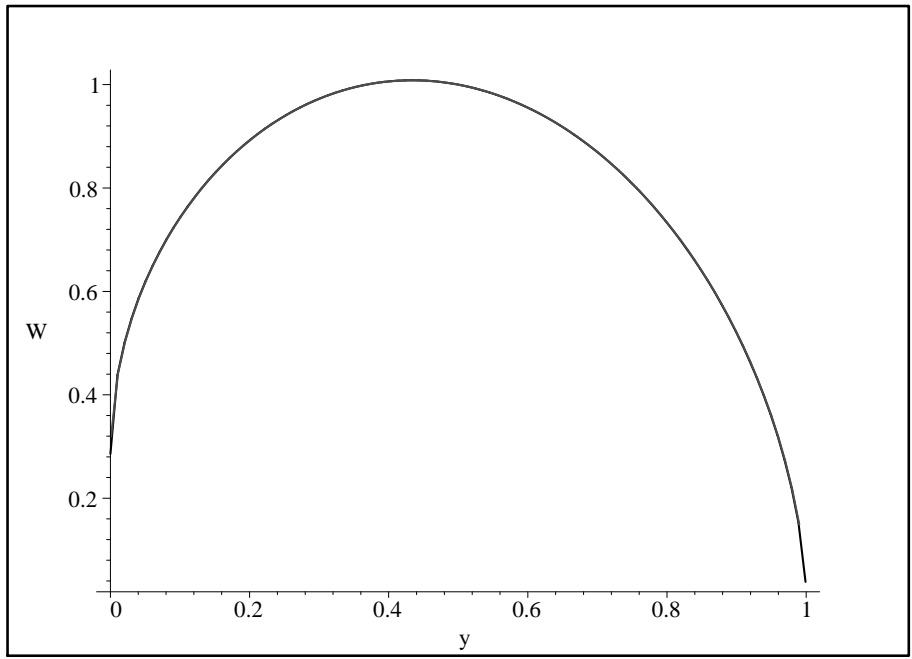}}
\vspace{0cm} \caption{A plot of $W(y)$, as a function of $y$, for
$\Lambda=1$, $k=1$, and $\Omega^2=-1.075...$ ($\Omega=1.037...i$).}
\end{figure}

\begin{figure}[H]
\centerline{\includegraphics[height=12cm,angle=-90]{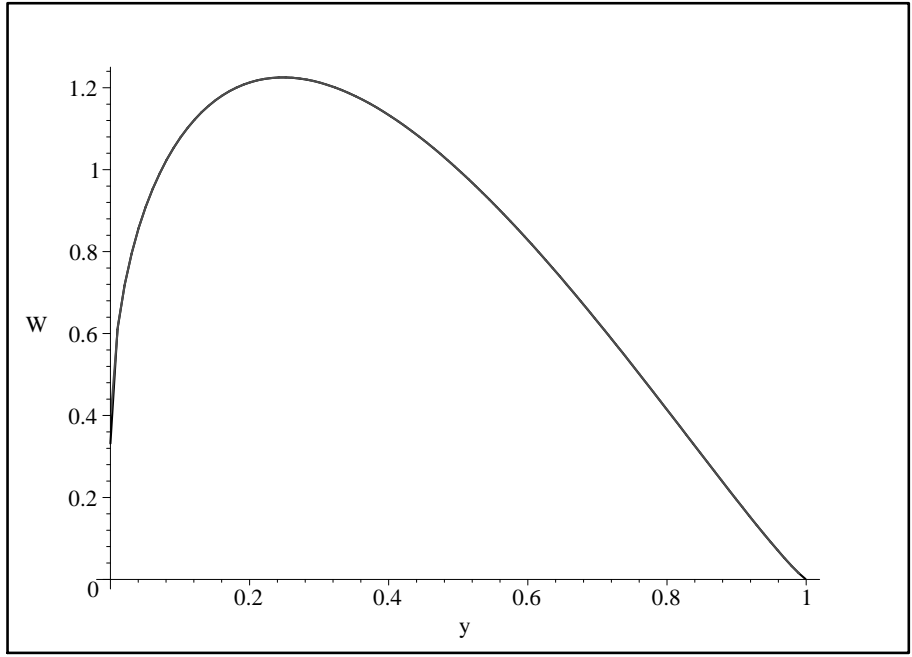}}
\vspace{0cm} \caption{A plot of $W(y)$, as a function of $y$, for
$\Lambda=1$, $k=2$, and $\Omega^2=-1.52...$ ($\Omega=1.23...i$).}
\end{figure}

Going back again to Figures 4 and 5, they correspond in each case to
the lowest eigenvalue $\Omega^2$, which, as shown, is negative.
Since they show the existence of solutions with $\Omega^2 <0$, we
conclude that, for $\kappa=1$, we do have unstable solutions that
are part of a complete spectrum, and therefore, in this case, i.e.,
$\kappa=1$, (and, as indicated for $\kappa =0$), the space time is
also linearly unstable.

\section{The case $k=0$. ``Radial'' perturbations.}

If we set $k=0$, $\ell=0$, the perturbations depend only on $t$, and
the ``radial'' variable $y$. We may analyze this case as the limit
as $k \to 0$ of the ``diagonal'' perturbations of the previous
Sections, but it is simpler instead to consider it directly as
perturbations that depend only on $y$ and $t$. In this case we may
choose a gauge where only $F_1(y)$, $F_2(y)$, and $F_3(y)$ are non
vanishing. This gauge is unique up to an additive constant in
$F_1(y)$. The resulting perturbation equations may be written in the
form:
\begin{eqnarray}
\label{ke001}
 \frac{d F_1}{dy} &=& {\frac { \left( 8 {\Lambda}^{2}{y}^{2}
 {\mu}^{2}-12 \mu \Lambda  \left( \mu-1-\kappa
 \right) y-9 \mu-9 \kappa \right)   }
 { \left( 4\,\Lambda\,y\mu-3 \right) ^{2}}}
{\frac {d F_3}{dy}} \nonumber \\
&& + \frac{2  \mu\,{\Omega}^{2}  {y}^{{\frac {1-\mu}{\mu}}}}{
  \left( 1-\Lambda\,y \right) ^{{\frac {3-\mu
}{3 \mu}}} \left( 4\,\Lambda\,y\mu-3 \right)} F_3 +{ \frac {12
\kappa\, \left( 2+\kappa \right) \Lambda\,\mu}{ \left(
4\,\Lambda\,y\mu-3 \right) ^{2}}} F_3, \\
  F_{{2}} & = &{\frac {6 y \left(\Lambda y-1 \right) \mu  }{4\,\Lambda\,y
\mu-3}} {\frac {dF_3}{dy}} +{\frac { 3 \kappa\, \left( 2+\kappa
 \right) }{4\,\Lambda\,y\mu-3}} F_3 \nonumber
\end{eqnarray}
and,
\begin{eqnarray}
\label{ke003}
 {\frac {d^{2}F_3}{d{y}^{2}}}& =
 &{\frac { \left( 16 {\Lambda}^{3}{\mu}^{2}{y}^{3}-8
 {\Lambda}^{2}\mu
 \left( 3+\mu \right) {y}^{2}-6 \Lambda \left(2 \mu \kappa -3\,\kappa-\mu
 -3 \right) y-9 \kappa-9 \right)   }{y \left( 4\,\Lambda\,y\mu-3
\right)  \left(1- \Lambda\,y \right)  \left(
-3-3\,\kappa+2\,\Lambda\,y\mu \right) }} \frac{dF_3}{dy}\nonumber \\
&&- \frac{ y^{\frac{1-2 \mu}{\mu}} \Omega^2}{ 3 \left( 1-\Lambda y
\right) ^{{ \frac {2 \mu+3}{3\mu}}}} F_3 +{ \frac {6 \left( 2+\kappa
\right) \left( 2 \kappa+1 \right)  \kappa \Lambda}{y \left( 4
\Lambda y\mu-3
 \right)  \left(1-\Lambda y \right)  \left(  2 \Lambda
 y\mu-3-3 \kappa \right) }} F_3
\end{eqnarray}

In accordance with (\ref{ke001}), this implies that both $F_2$, and
$F_1$, (up to a constant) are given, once we find the solutions for
(\ref{ke003}). However, instead of $F_3(y)$, it will be simpler here
to consider again the gauge invariant $W$, which, for $F_4=0$,
reduces to,
\begin{equation}
\label{ke005}
    W(y)= (3+3\kappa-2 \Lambda \mu y) F_3(y)
\end{equation}

Replacing in (\ref{ke003}), we find,
\begin{equation}
\label{ke003a}
 {\frac {d^{2}W}{d{y}^{2}}}  = {\frac {
\left( 4 \Lambda y\mu-6 \Lambda y+3 \right) }{y \left( 1-\Lambda y
\right) \left( 4\Lambda y\mu-3 \right) }}\frac{dW}{dy}
 -\frac{{y}^{{\frac
{1-2 \mu}{\mu}}} \Omega^2 }{3 \left( 1- \Lambda y \right) ^{{\frac
{2 \mu+3}{3\mu}}}} W
 -{\frac { 2 \Lambda \left( 2 \mu-3 \right)  }
 {y \left( 1-\Lambda y \right)  \left( 4\Lambda y\mu-3
 \right) }} W
\end{equation}
which is just the limit $k \to 0$ of (\ref{diag40}). We may,
therefore, use the results of Section V and VI to analyze the
resulting spectrum of $\Omega$. We notice here that if we rescale
$y$ to $y/\Lambda$, and $\Omega^2$ to $\Omega^2/\Lambda^{1/\mu}$,
(\ref{ke003a}) takes the form,
\begin{equation}
\label{ke003b}
 {\frac {d^{2}W}{d{y}^{2}}}  = {\frac {
\left( 4  y\mu-6  y+3 \right) }{y \left( 1- y \right) \left( 4
y\mu-3 \right) }}\frac{dW}{dy}
 -\frac{{y}^{{\frac
{1-2 \mu}{\mu}}} \Omega^2 }{3 \left( 1-  y \right) ^{{\frac {2
\mu+3}{3\mu}}}} W
 -{\frac { 2 \left( 2 \mu-3 \right)  }{y \left( 1- y \right)
  \left( 4 y\mu-3
 \right) }} W
\end{equation}
and, therefore, in what follows we will set $\Lambda=1$, and,
without loss of generality, analyze directly (\ref{ke003b}). There
remains the problem of determining the spectrum of $\Omega$. For
this purpose we simply consider again the derivations of Sections
III, and Section IV, (in the limit $k \to 0$), noticing that both
$r(y)$ and $K(y)$, are independent of $k$, and depend only on $\mu$.
Regarding the construction of $\widetilde{V}(r)$, we notice that in
this case, for $\Omega=0$, eq. (\ref{ke003b}) has the exact
solution, \cite{explain}
\begin{equation}\label{ke018}
   W(y)=C_1\left[3+(4\mu-6)y\right]+C_2\left[(3+(4\mu-6)y)
   \ln\left(y/(1-y)\right)
   +6-8\mu y\right]
\end{equation}
where $C_1$, and $C_2$ are arbitrary constants. We, therefore, set,
\begin{equation}\label{ke019}
   W^{(0)}(y)=C_1\left[3+(4\mu-6)y\right]
\end{equation}
and, replacing appropriately in (\ref{sp24}), we find,
\begin{eqnarray}
 \label{ke027}
    \widetilde{V}(r)&=& \frac {(1-y)^{1/\mu-4/3}}{16 {y}^{{1/\mu}}
    \left( 4\,\mu\,y
-6\,y+3 \right) ^{2}{\mu}^{2}} \left[ 81- \left( 864 \mu+1152
{\mu}^{3}+324-2160 {\mu}^ {2} \right) y  \right. \\ && \left. +
\left(324 -6480 {\mu}^{2}+3024 \mu+3072 {\mu}^{3}
 \right) {y}^{2} + 32 \mu \left( 2 \mu-3 \right)  \left( 4 {\mu}^{2}-36
\mu+27
 \right) {y}^{3} \right] \nonumber
\end{eqnarray}
The term in square brackets in (\ref{ke027}) is a polynomial in $y$
and therefore it is bounded above and below in $0\leq y \leq 1$. The
factor in front of this polynomial is positive definite and
continuous in $0 < y < 1$ and diverges to $+\infty$, both at $y=0$
and $y=1$. The potential $ \widetilde{V}(r)$ is therefore continuous
and bounded from below in $0 < y < 1$. To analyze (\ref{ke027})
further we notice that near $y=0$, to leading order, we have,
\begin{equation}
 \label{ke029}
 \widetilde{V}(r) = \frac{9}{16 \mu^2 y^{1/\mu}}+\dots
\end{equation}
while, in accordance with (\ref{diag56}), again to leading order, we
have,
\begin{equation}
 \label{ke031}
 \frac{1}{y^{1/\mu}} = \frac{4 \mu^2}{3 r^2} +\dots
\end{equation}

Therefore, also to leading order as $y \to 0$, we have,
\begin{equation}
 \label{ke033}
 \widetilde{V}(r) = \frac{3}{4 r^2}+\dots
\end{equation}
and this implies that near $y=0$, also to leading order, the general
solution of (\ref{sp08}) is, in this case, of the form,
\begin{equation}
 \label{ke035}
    \chi(r) = C_1 r^{3/2} +C_2 r^{-1/2} + \dots
\end{equation}
where $C_1$ and $C_2$ are arbitrary constants.

Similarly, near $y=1$, we have $r \to r_1$, and, to leading order,
\begin{equation}
 \label{ke037}
    (1-y)^{\frac{3-4 \mu}{3\mu}}= \frac{12 \mu^2}{(4\mu-3)^2
    (r_1-r)^2}+\dots
\end{equation}
and, therefore, again near $y=1$,
\begin{equation}
 \label{ke039}
 \widetilde{V}(r) = \frac{3}{4 (r_1-r)^2}+\dots
\end{equation}
and this implies that near $r=r_1$, the general solution $\chi(r)$
of (\ref{sp08}), to leading order, behaves as,
\begin{equation}
 \label{ke041}
    \chi(r) = C_3 (r_1-r)^{3/2} +C_4 (r_1-r)^{-1/2} + \dots
\end{equation}

These results imply that we will obtain a self adjoint extension of
(\ref{sp08}) if we impose $C_2=0$, and $C_4=0$, as boundary
conditions on its solutions. The relation of the solutions of the
self adjoint extension of (\ref{sp08}) to the general solution of
the evolution of arbitrary perturbations follows the lines indicated
in the previous Sections, and we shall not repeat it here.

\begin{figure}[H]
\centerline{\includegraphics[height=12cm,angle=-90]{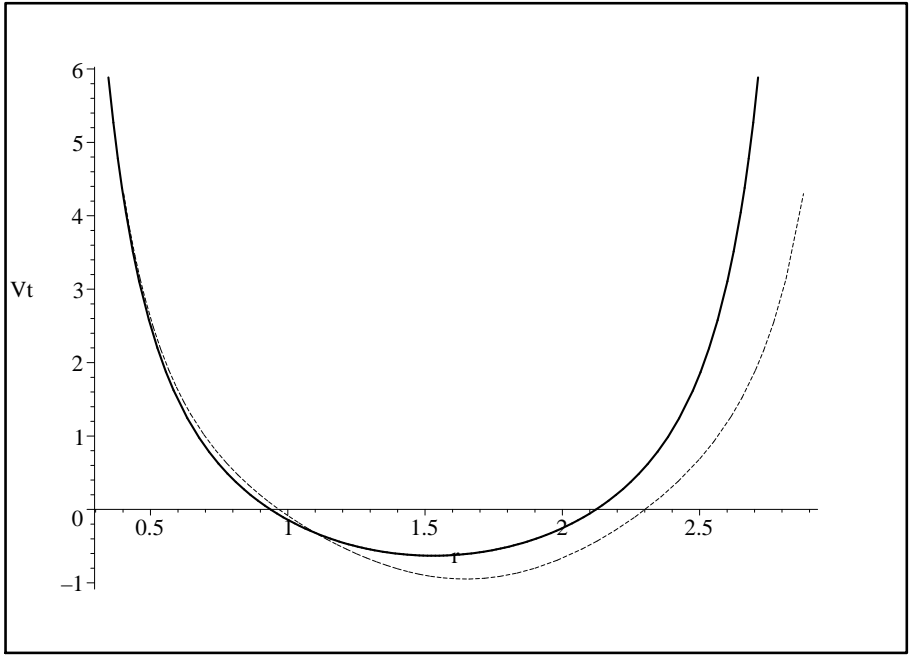}}
\vspace{0cm} \caption{A plot of $\widetilde{V}(y)$, (Eq.
(\ref{ke027})), as a function of $r$, for $\Lambda=1$, and for
$\mu=3/2$, (solid line), and $\mu=6/5$ (dashed line).}
\end{figure}

There remains the problem of analyzing the spectrum of this self
adjoint extension. From the general properties of
$\widetilde{V}(r)$, this spectrum will be discrete and bounded from
below. However, in general, and as shown in Figure 4 for two
examples, $\widetilde{V}(r)$ is not positive definite. Therefore,
the lowest eigenvalue of (\ref{sp08}) might be negative. This can
only be analyzed numerically. Here we have the simplifying feature
that $\widetilde{V}(r)$ is given explicitly as a function of $y$. If
we define $\chi_1(y)= \chi(r(y))$, and change coordinates in
(\ref{sp08}) back to $y$ we find,
\begin{eqnarray}
\label{ke043}
 {\frac {d^{2}\chi_1}{d{y}^{2}}} &= & {\frac {
 \left( 8 \mu y+3-6 \mu \right)   }{ 6y \left( 1-y \right) \mu}}
 \frac{d \chi_1}{dy}
 - \frac{ y^{{
\frac {1-2\mu}{\mu}}}{\Omega}^{2}}{3\left( 1-y \right) ^{
{\frac {2 \mu+3}{3\mu}}}} \chi_1 \\
 & &-\frac{1}{48 {y}^{2} \left( 1-y \right) ^{2}
 \left( 4 \mu y-6 y+3 \right) ^{2}{\mu}^{2}}
  \left[ 81+32 \mu  \left( 2 \mu-3 \right)  \left( 4 {\mu}
^{2}-36 \mu+27 \right) {y}^{3} \right. \nonumber \\
 & & \left.+ \left( 3024
\mu+3072 {\mu}^{3}+324- 6480 {\mu}^{2} \right) {y}^{2}- \left(
 864 \mu+1152 {\mu}^{3}+324 -2160 {\mu}^{2} \right) y \right]
\chi_1 \nonumber
\end{eqnarray}

The boundary conditions for (\ref{ke043}) corresponding to the self
adjoint extension of (\ref{sp08}) are in this case,
\begin{eqnarray}
\label{ke045}
  \chi_1(y) & \to & A \,y^{\frac{3}{4\mu}} \;\;;\;\; {\mbox{for}}
  \;\; y \to 0  \\
  \chi_1(y) & \to & B \,(1-y)^{\frac{4\mu-3}{4\mu}} \;\;;\;\;
  {\mbox{for}}\;\; y \to
  1 \nonumber
\end{eqnarray}
where $A$ and $B$ are constants. The range of $\mu$ we are
interested in is $1\leq \mu\leq 3$. However, if we take a solution
$\chi_1(y)$ of (\ref{ke043}), corresponding to the eigenvalue
$\Omega^2$, and we define,
\begin{equation}\label{ke047}
    \chi_2(y)=\chi_1(1-y)
\end{equation}
one can check that $\chi_2(y)$ is also a solution of (\ref{ke043})
for the same $\Omega^2$, but with $\mu$ replaced by,
\begin{equation}\label{ke049}
    \nu=\frac{3 \mu}{4 \mu-3}
\end{equation}
and satisfying also (\ref{ke045}), with the same replacement. Since
for $1\leq \mu \leq 3/2$, we have $3 \geq \nu\geq 3/2$, we only need
to study the range $1\leq \mu \leq 3/2$. We notice that for
$\mu=1\;,\;(\kappa=0)$, and $\Omega=0$, we have the exact solution,
\begin{equation}\label{ke051}
    \chi_1(y)=C_1\frac{y^{\frac{3}{4}} (1-y)^{\frac{1}{4}}}{2y-3}
    +C_2 \frac{4y-3}{y^{\frac{1}{4}} (1-y)^{\frac{1}{12}} (2 y -3)}
\end{equation}
where $C_1$ and $C_2$ are constants. Clearly, the solution
corresponding to the self adjoint extension is obtained by setting
$C_2=0$. Since in this case (\ref{ke051}) has no nodes, it must
correspond to the lowest eigenvalue. Similarly, for $\mu=3$, we have
the exact solution,
\begin{equation}\label{ke053}
    \chi_1(y)=C_1\frac{(1-y)^{\frac{3}{4}} y^{\frac{1}{4}}}{2y+1}
    +C_2 \frac{4 y-1}{(2y+1)(1-y)^{\frac{1}{4}} y^{\frac{1}{12}}}
\end{equation}
corresponding again to the eigenvalue $\Omega^2=0$, in agreement
with our previous discussion. As a check for the numerical
integration procedure, we reobtained numerically the solution
(\ref{ke051}), (with $C_2=0$). For the next higher eigenvalue we
obtained $\Omega=1.635...$, ($\omega^2=2.673...$).

We have also explored numerically the range $1< \mu\leq 3/2$, and
found that the lowest eigenvalue $\Omega^2$ was in all cases larger
than zero. For instance, for $\mu=6/5$ we found $\Omega=0.854...$,
and for $\mu=3/2$ we obtained $\Omega=1.05...$. For the next
eigenvalue we found $\Omega=2.17...$, for $\mu=6/5$, and
$\Omega=2.35...$, for $\mu=3/2$.

The general conclusion here is that there are no unstable modes for
the ``radial'' perturbations ($k=0$, $\ell=0$). We recall that
solutions with $\Omega=0$, for $\chi(r)$, do not correspond to
solutions for $W(y)$ with the same eigenvalue.

\section{Summary and conclusions.}

In this paper we have analyzed the linear stability of the Linet -
Tian space times with a positive cosmological constant $\Lambda$.
These space times contain a particular symmetry relating their two
space like Killing vectors that is not present in the case of a
negative $\Lambda$. This symmetry allows for a simplification
regarding the parameter space to be analyzed. In a separate study in
\cite{glei3} it was shown that a particular class of perturbations
can be restricted to the ``diagonal'' form (\ref{gau07}). This leads
to a consistent set of perturbation equations, but it still contain
a gauge ambiguity that cannot be removed. To deal with this problem
we introduced a gauge invariant function of the metric
perturbations, called $W(y)$ in the text, that satisfies an equation
that takes the form of a linear boundary value problem, from which
one can extract the allowed values of $\Omega$, the frequency of the
perturbation modes. Although these solutions are well defined, it is
not at all clear if, and in what sense, the set of solutions for
$W(y)$ is complete, and in what way they are related to the
evolution of arbitrary initial data. This problem is solved by the
introduction, using the Darboux transformation, of a related self
adjoint problem, with a complete set of eigenvalues and
eigenfunctions. This leads to a definite form for the expansion of
arbitrary initial data in terms of the modes of $W(y)$, and its
corresponding eigenvalues $\Omega$. The remaining problem, the
existence of solutions of the equation for $W(y)$ with the required
properties is solved numerically, and we display several examples of
those solutions. This in turn provides a complete proof of the
linear instability of the Linet - Tian metric. It turns out that for
$\kappa=0$, and $\kappa=1$ the perturbation equation require a
special treatment that is also given in this paper, again showing
that even for those cases the space time is linearly unstable. We
have also included a discussion of ``radial'' perturbations, i.e.,
those preserving the space like Killing vectors. We find that in
this case there are no unstable modes. In conclusion we have shown
that linear perturbations of the Linet - Tian metric contain
unstable modes, and that these unstable modes are part of a complete
set of solutions, and therefore, the linear instability is a generic
feature of these metrics.

\section*{Acknowledgments}

I am grateful to G. Dotti for many helpful comments, suggestions and
criticisms.

\appendix

\section{Regularity of $W$ at $y=y_s$.}

Consider a function $W(y)$ that is a solution, in a certain range
$y_1 \leq y \leq y_2$, of an equation of the general form,
\begin{equation}\label{A02}
    -\frac{d^2 W}{dy^2}+Q_1\frac{d W}{dy} +Q_2 W
    =\Omega^2 Q_3
\end{equation}
where $Q_1$, $Q_2$, and $Q_3$ are functions of $y$. We assume no
particular boundary conditions on either $W$ or the functions $Q_i$,
but we shall assume that in a neighbourhood of a point $y=y_s$, with
$y_s$ interior to the interval $(y_1,y_2)$, the functions $Q_i$
admit Laurent expansions of the form,
\begin{eqnarray}
\label{A04}
  Q_1(y) &=& \frac{2}{y-y_s}+a_1+a_2(y-y_s)+a_3(y-y_s)^2+a_4(y-y_s)^3+\dots \nonumber \\
  Q_2(y) &=& \frac{b_0}{y-y_s}+b_1+b_2(y-y_s)+b_3(y-y_s)^2+b_4(y-y_s)^3+\dots  \\
  Q_3(y) &=& c_1+c_2(y-y_s)+c_3(y-y_s)^2+c_4(y-y_s)^3+\dots \nonumber
\end{eqnarray}
where dots indicate higher order terms. We also assume $c_1 > 0$.
Then, if the coefficients in (\ref{A04}) satisfy the relations,
\begin{eqnarray}
\label{A06}
  c_2 &=& c_1(a_1+b_0) \\
  b_2 &=& b_{{1}}a_{{1}}-\frac{1}{2}{a_{{1}}}^{2}b_{{0}}-\frac{3}{4}a_{{1}}{b_{{0}}}^{2}
  +\frac{1}{2}a_{{2}}b_{{0}}+b_{{1}}b_{{0}}-\frac{1}{4}{b_{{0}}}^{3} \nonumber
\end{eqnarray}
a straightforward computation shows that the general solution of
(\ref{A02}) takes the form,
\begin{equation}
\label{A08}
    W(y) = w_0+w_1
    (y-y_s)+w_2(y-y_s)^2+w_3(y-y_s)^3+w_4(y-y_s)^4+\dots
\end{equation}
where $w_0$, and $w_3$ are arbitrary constants,
\begin{eqnarray}
\label{A10}
  w_1 &=& -\frac{1}{2}b_0 w_0  \\
  w_2 &=& \frac{1}{4}\left(b_0^2-2b_1+a_1b_0+2 c_1 \Omega^2\right)
  w_0 \nonumber
\end{eqnarray}
and $w_4$, and higher order terms are (homogeneous) linear functions
of $w_0$, and $w_3$. This implies that the general solution of
(\ref{A02}), with the $Q_i$ satisfying (\ref{A06}), independently of
any boundary conditions at either $y=y_1$, or $y=y_2$, is a linear
combination of two regular solutions, one that near $y=y_s$ behaves
as,
\begin{equation}
\label{A08a}
    W(y) =
    w_0\left(1+\widetilde{w}_1(y-y_s)+\widetilde{w}_2(y-y_s)^2+\widetilde{w}_3(y-y_s)^3+\dots
    \right)
\end{equation}
and another solution that near $y=y_s$ behaves as,
\begin{equation}
\label{A08b}
    W(y) = w_3\left[(y-y_s)^3+\bar{w}_4(y-y_s)^4+\dots \right]
\end{equation}

\section{Regularity of $\widetilde{V}(r)$ at $y=y_s$.}

We first write (\ref{sp24}) in the form,
\begin{eqnarray}
\label{B02}
  \widetilde{V}(r(y)) &=&  -\frac{1}{4 Q_3^3}\left[ Q_3 \frac{d^2 Q_3}{dy^2}
  -\frac{5}{4}\left(\frac{d Q_3}{dy}\right)^2\right]
   -\frac{1}{4 Q_3}\left(Q_1^2+4 Q_2-2 \frac{dQ_1}{dy}\right)  \\
   & & + 2 \Omega_0^2 +\frac{2}{Q_3}\left(\frac{1}{W^{(0)}_1}
   \frac{dW^{(0)}_1}{dy} -\frac{1}{2}Q_1+\frac{1}{4 Q_3}\frac{dQ_3}{dy}\right)^2
   \nonumber
\end{eqnarray}

Using now the results of Appendix A, where we assume for
${W^{(0)}_1}$ the general expansion (\ref{A08}), we find,
\begin{eqnarray}
\label{B04}
  \widetilde{V}(r(y)) &=&   {\frac { \left( 24 a_{{2}}+19 {a_{{1}}}^{2}+
14 a_{{1}}b_{{0}}+7 {b_{{0}}}^{2}+48 b_{{1}} \right) c_{{1}}-40 c_
{{3}}}{16 {c_{{1}}}^{2}}} \\
& &-2 {\Omega}_0^{2}+ \mathcal{O}\left(y-y_s\right) \nonumber
\end{eqnarray}

Then, since $dr/dy=\sqrt{Q_3}$, near $y=y_s$ we have,
\begin{equation}
\label{B06}
    y=y_s+\frac{1}{\sqrt{c_1}}(r-r_s)+\mathcal{O}\left(r-r_s\right)
\end{equation}
where $r_s=r(y_s)$, and this implies that near $r-r_s$ we have,
\begin{eqnarray}
\label{B08}
  \widetilde{V}(r) &=&   {\frac { \left( 24 a_{{2}}+19 {a_{{1}}}^{2}+
14 a_{{1}}b_{{0}}+7 {b_{{0}}}^{2}+48 b_{{1}} \right) c_{{1}}-40 c_
{{3}}}{16 {c_{{1}}}^{2}}} \\
& &-2 {\Omega}_0^{2}+ \mathcal{O}\left(r-r_s\right) \nonumber
\end{eqnarray}
and therefore $\widetilde{V}(r)$ is regular at $r=r_s$, irrespective
of the choice of $W^{(0)}_1(y)$, provided only that $W^{(0)}_1(y_s)
\neq 0$.

\section{The self adjoint problem associated to $W$.}

We consider again (\ref{B02}). As shown in Appendix B,
$\widetilde{V}(y(r))$ is regular at $y=y_s$ for any solution
$W^{(0)}_1(y)$ of (\ref{diag40}) that does not vanish at $y=y_s$.
Let us now assume that $W^{(0)}_1$ satisfies the boundary conditions
(\ref{diag26},\ref{diag28}), and that it does not vanish anywhere in
$0\leq y \leq 1/\Lambda$. Then, since $Q_3(y) > 0$ in $0 < y <
1/\Lambda$, and $Q_1(y)$, and $Q_2(y)$ are regular in that interval,
away from $y=y_s$, $\widetilde{V}(y(r))$ will also be regular in
that interval. As we approach $y=0$, to leading terms, we have,
\begin{eqnarray}
\label{sad04}
  Q_1(y) & \simeq & -\frac{1}{y}-\frac{(8 \mu-3)\Lambda}{3} \nonumber \\
  Q_2(y) & \simeq & -\frac{k^2(2\kappa+3)(2\kappa+1)y^{\frac{\kappa}{\mu}}}{3 y}
   +\frac{2 \Lambda(2\mu-3)}{3 y} \\
  Q_3(y) & \simeq & \frac{y^{\frac{1}{\mu}}}{3 y^2}
  +\frac{(3+2\mu)\Lambda y^{\frac{1}{\mu}}}{9\mu y} \nonumber
\end{eqnarray}
and, therefore, we have,
\begin{equation}
\label{sad06}
    \widetilde{V}(y(r))= \frac{9}{16 \mu^2 y^{\frac{1}{\mu}}}+
    \mathcal{O}\left(y^0\right)
\end{equation}
and, since near $y=0$ we have,
\begin{equation}
\label{sad08}
    r(y)\simeq \frac{2 \mu y^{\frac{1}{2\mu}}}{\sqrt{3}}
\end{equation}
we finally get that as $r \to 0$, we have,
\begin{equation}
\label{sad10}
    \widetilde{V}(r)= \frac{3}{4 r^2}+ \dots
\end{equation}
where dots indicate higher order terms.

Similarly, near $y=1/\Lambda$ we have,
\begin{eqnarray}
\label{sad12}
  Q_1(y) & \simeq & -\frac{\Lambda}{1-\Lambda y}+\dots \nonumber \\
  Q_2(y) & \simeq & \frac{(4 \kappa-1)\Lambda^{\frac{\mu-\kappa}{\mu}}
  k^2}{(2\kappa+1)^2} (1-\Lambda y)^{-\frac{3 \kappa+5 \mu}{3 \mu}} +\dots \\
  Q_3(y) & \simeq & \frac{\Lambda^{\frac{2 \mu -1}{\mu}}}{3}
  (1-\Lambda y)^{-\frac{3+2\mu}{\mu}} +\dots \nonumber
\end{eqnarray}
then, replacing in (\ref{B02}) we find,
\begin{equation}
\label{sad16}
    \widetilde{V}(y(r))= \frac{(4\mu-3)^2
    \Lambda^{\frac{1}{\mu}}}{16 \mu^2} (1-\Lambda
    y)^{\frac{3-4\mu}{3\mu}} +\dots
\end{equation}
Using now (\ref{diag48}), near $y=1/\Lambda$ we have,
\begin{equation}
\label{sad18}
    r(y)= r_1 - \frac{2 \sqrt{3} \mu}{\Lambda^{\frac{1}{\mu}}
    (4\mu-3)} (1-\Lambda y)^{\frac{4\mu-3}{6 \mu}} +\dots
\end{equation}
where $r_1=r(1/\Lambda)$, and this implies,
\begin{equation}
\label{sad20}
    \widetilde{V}(r)= \frac{3}{4 (r_1-r)^2} +\dots
\end{equation}
where in all these expressions dots indicate higher order terms.

Therefore, if $W(y)$ does not vanish in $0 < y< 1/\Lambda$, since
the $Q_i$ are bounded away from $y=0$, $y=y_s$ and $y =1/\Lambda$,
and $\widetilde{V}(r)$, as already shown, is regular at $y=y_s$, we
have that $\widetilde{V}(r)$ will have a lower bound in $0 < r <
r_1$, diverging at $r=0$ and at $r=r_1$ as given by (\ref{sad10}),
and (\ref{sad20}), respectively. But this, in turn, implies that
near $r=0$ the solutions $\chi(r)$ of (\ref{sp08}) behave as,
\begin{equation}
\label{sad22}
    \chi(r) \simeq C_1 r^{\frac{3}{2}}+C_2 r^{-\frac{1}{2}}
\end{equation}
and near $r=r_1$ as,
\begin{equation}
\label{sad24}
    \chi(r) \simeq C_3 (r_1-r)^{\frac{3}{2}}+C_4(r_1- r)^{-\frac{1}{2}}
\end{equation}
where the $C_i$ are arbitrary constants, and, therefore, we may
construct a self adjoint extension of (\ref{sp08}) by imposing the
boundary conditions that $C_2=0$, and $C_4=0$. The resulting
spectrum of $\Omega^2$ will be fully discrete and bounded from
below.

\section{Gauge transformations of the perturbed Linet-Tian metric.}

Consider the perturbation expansion (\ref{gau01}), restricted, for
simplicity, to functions of the form (\ref{gau03}), but with
$\ell=0$. We write it in the form,
\begin{eqnarray}
\label{gt01}
  ds^2 &=& -\frac{y^{\frac{1}{3}+\frac{p_1}{2}}}{(1-\Lambda y)^{\frac{p_1}{2}-\frac{1}{3}}}
  \left(1 +\epsilon e^{{i(\Omega t - k z)}} F_1\right) dt^2
 +\frac{1}{3 y(1-\Lambda y)}
  \left(1 +\epsilon e^{{i(\Omega t - k z)}} F_2\right)
  dy^2 \nonumber
  \\
   &+& \frac{y^{\frac{1}{3}+\frac{p_2}{2}}}{(1-\Lambda y)^{\frac{p_2}{2}-\frac{1}{3}}}
  \left(1 +\epsilon e^{{i(\Omega t - k z)}} F_3\right) dz^2
 + \frac{y^{\frac{1}{3}+\frac{p_3}{2}}}{(1-\Lambda y)^{\frac{p_3}{2}-\frac{1}{3}}}
  \left(1 +\epsilon e^{{i(\Omega t - k z)}} F_4\right) d\phi^2  \\
   &+& 2\epsilon e^{{i(\Omega t - k z)}} \left( F_5 dt dy+F_6 dt
   dz+F_7 dz dy+F_8 dt d\phi+F_9 dy d\phi+F_{10} dz d\phi \right)
   \nonumber
\end{eqnarray}
where $F_i=F_i(y)$, and $\epsilon$ is used to keep track of the
perturbation order. Under a first order transformation to new
coordinates $(T,Y,Z,\Phi)$, such that,
\begin{eqnarray}
\label{gt03}
  t &=& T + \epsilon e^{i(\Omega T - k Z)} Q_1(Y) \;\;
  ; \;\; y = Y + \epsilon e^{i(\Omega T - k Z)} Q_2(Y)  \nonumber \\
  z &=& Z + \epsilon e^{i(\Omega T - k Z)} Q_3(Y) \;\;
  ; \;\; \phi = \Phi + \epsilon e^{i(\Omega T - k Z)} Q_4(Y)
\end{eqnarray}
where the functions $Q_i$ are arbitrary, the form of the metric
(\ref{gt01}) is preserved, but the coefficients $F_i(y)$ are changed
to $\widetilde{F}_i(y)$, where \cite{y2Y},
\begin{eqnarray}
\label{gt05}
  \widetilde{F}_1(y) &=& F_1(y) + 2 i \Omega Q_1(y)
     -\frac{4 \Lambda y -2 -3 p_1}{ 6 y (1-\Lambda y)} Q_1(y) \nonumber \\
  \widetilde{F}_2(y) &=& F_2(y)
  -\frac{1-2\Lambda y }{y(1-\Lambda y)}Q_2(y) +2 \frac{dQ_2}{dy}  \nonumber \\
  \widetilde{F}_3(y) &=& F_3(y)
  +\frac{2+3 p_2 -4 \Lambda y}{6 y (1-\Lambda y)} Q_2(y) -2i k Q_3(y)   \nonumber \\
  \widetilde{F}_4(y) &=& F_4(y)
  +\frac{2+3 p_3-4\Lambda y}{6 y (1-\Lambda y)} Q_2(y) \nonumber \\
  \widetilde{F}_5(y) &=& F_5(y)
  +\frac{i \Omega}{3 y (1-\Lambda y)} Q_2(y)- y^{\frac{1}{3}+
\frac{p_1}{2}}(1-\Lambda y)^{\frac{1}{3}-\frac{p_1}{2}}\frac{dQ_1}{dy}\nonumber \\
  \widetilde{F}_6(y) &=& F_6(y)
  +i \Omega y^{\frac{1}{3}+
\frac{p_2}{2}}(1-\Lambda y)^{\frac{1}{3}-\frac{p_2}{2}} Q_3(y)
 +i k y^{\frac{1}{3}+
\frac{p_1}{2}}(1-\Lambda y)^{\frac{1}{3}-\frac{p_1}{2}} Q_1(y) \nonumber \\
  \widetilde{F}_7(y) &=& F_7(y)
  -\frac{i k}{3 y (1-\Lambda y)}Q_2(y)
   +y^{\frac{1}{3}+
\frac{p_2}{2}}(1-\Lambda
y)^{\frac{1}{3}-\frac{p_2}{2}}\frac{dQ_3}{dy}
\end{eqnarray}
and,
\begin{eqnarray}
\label{gt07}
  \widetilde{F}_8(y) &=& F_8(y)
  +i \Omega y^{\frac{1}{3}+
\frac{p_3}{2}}(1-\Lambda y)^{\frac{1}{3}-\frac{p_3}{2}} Q_4(y)\nonumber \\
  \widetilde{F}_9(y) &=& F_9(y)
  +y^{\frac{1}{3}+
\frac{p_3}{2}}(1-\Lambda y)^{\frac{1}{3}-\frac{p_3}{2}} \frac{dQ_4}{dy}\nonumber \\
  \widetilde{F}_{10}(y) &=& F_{10}(y)
  -i k y^{\frac{1}{3}+
\frac{p_3}{2}}(1-\Lambda y)^{\frac{1}{3}-\frac{p_3}{2}} Q_4(y)
\end{eqnarray}

This implies that the sets
$\mathcal{S}_1=\{F_1,F_2,F_3,F_4,F_5,F_6,F_7\}$, and
$\mathcal{S}_2=\{F_8,F_9,F_{10}\}$ transform independently of each
other. This is reflected also in the perturbation equations, since,
as can be checked, they separate also into two set, one coupling
only the functions in $\mathcal{S}_1$, and the other only those in
$\mathcal{S}_2$.

Going back to (\ref{gt07}), we can, e.g., choose $Q_4$ such
$F_8(y)=0$, and that fixes the gauge, in the sense that $F_9(y)$,
and $F_{10}(y)$ are uniquely determined \cite{nondiagonal}. In the
case of (\ref{gt05}) we may choose the $Q_i$ such that only $F_1$,
$F_2$, $F_3$, and $F_4$, (the ``diagonal'' terms), are non
vanishing, and one can check that this choice leads to a consistent
set of perturbation equations (given by (\ref{gau07})) in Section
III. This restriction, however, does not determine completely these
functions, because, as can be checked, a transformation with the set
of functions,
\begin{eqnarray}
\label{gt09}
  Q_1(y) &=& \Omega y^{\frac{p_2}{4}-\frac{p_1}{4}}
  (1-\Lambda y)^{\frac{p_1}{4}-\frac{p_2}{4}} Q_0 \nonumber \\
  Q_2(y) &=& \frac{3}{4} i (p_1-p_2) y^{\frac{1}{3}+\frac{p_1}{4}+\frac{p_2}{4}}
   (1-\Lambda y)^{\frac{1}{3}-\frac{p_1}{4}-\frac{p_2}{4}} Q_0 \\
  Q_3(y) &=& - k y^{\frac{p_1}{4}-\frac{p_2}{4}}
  (1-\Lambda y)^{\frac{p_2}{4}-\frac{p_1}{4}} Q_0 \nonumber
\end{eqnarray}
where $Q_0$ is an arbitrary function of $\Omega$, leaves the
``diagonal'' form invariant. This implies that to any solution of
the perturbation equations for $F_1$, $F_2$, $F_3$, and $F_4$, even
the trivial one where all $F_i=0$, we may add a solution of the form
(\ref{diag08}), and still have a solution.


\begin{thebibliography}{99}
\bibitem{conicity} One may also include ``conicity'' parameters, as
in \cite{griffiths}. They correspond to rescalings of the
corresponding coordinates, and play no role in the present analysis.
\bibitem{y2Y} Notice that to first order in $\epsilon$ we may
replace $(T,Y,Z,\Phi)$ by $(t,y,z\phi)$ in the equations that
follow.
\bibitem{nondiagonal} One can straighforwardly check that this choice of gauge leads
to a well determined self adjoint problem, with a complete set of
solutions with $\Omega^2 >0$. Therefore, this sector has only stable
modes, and will not be further analyzed here.
\end{thebibliography}
\end{document}